\documentstyle[prb,twocolumn,aps]{revtex}
\begin{document}
\draft
\title{Harmonic Solid Theory of Photoluminescence in the High Field 
Two-Dimensional Wigner Crystal}
\author{S.\ Kodiyalam$^\dagger$
, H.A.\ Fertig$^\ddagger$ 
and S.\ Das Sarma$^\dagger$}
\address{$^\dagger$Department of Physics, University of Maryland, College Park, 
Maryland 20742-4111}
\address{$^\ddagger$Department of Physics and Astronomy, University of 
Kentucky, Lexington, Kentucky 40506-0055}
\address{\mbox{ }}
\address{\parbox{16cm}{\rm \mbox{ }\mbox{ }\mbox{ }
Motivated by recent experiments on radiative recombination of two-dimensional 
electrons in acceptor doped GaAs-AlGaAs heterojunctions as well as the
success of a harmonic solid model in describing tunneling between 
two-dimensional electron systems, we calculate within the harmonic 
approximation and the time dependent perturbation theory the line 
shape of the 
photoluminescence spectrum corresponding to the recombination of an electron 
with a hole bound to an acceptor atom. The recombination process is modeled
as a sudden perturbation of the Hamiltonian for the in-plane degrees of 
freedom of the electron. We include in the 
perturbation, in addition to changes in the equilibrium positions of electrons,
 changes in the curvatures of the harmonically approximated potential.
The computed spectra have line shapes similar to that seen in a recent 
experiment. The spectral width, however, is roughly a factor of 3 smaller than 
that seen in experiment if one assumes a perfect Wigner crystal for the initial 
state state of the system, whereas a simple random disorder model yields a 
width a 
factor of 3 too large. We speculate on the possible mechanisms that may lead to
 better quantitative agreement with experiment.
}} 
\address{\mbox{ }}
\address{\mbox{ }}
\address{\parbox{16cm}{\rm PACS numbers: 73.20Dx ; 71.45.-d ;  78.20.Ls ;
73.40.Hm}}
\maketitle

\makeatletter
\global\@specialpagefalse
\def\@oddhead{REV\TeX{} 3.0\hfill Das Sarma Group Preprint, 1997}
\let\@evenhead\@oddhead
\makeatother
             
\narrowtext

\section{Introduction}
The properties of non-relativistic electrons in a jellium 
background have been a subject of experimental and theoretical interest. In its 
quantum mechanical description, the Planck's constant $\hbar$ and the 
electron charge $e$ and mass $m$, conveniently provide the natural scales  
for measuring the largeness/smallness of all other input parameters (such as 
the average density of electrons) that may enter the Hamiltonian of the 
system. Using the variational principle, Wigner \cite{W} pointed out that at 
low electron densities a (three dimensional) system of electrons with a 
wavefunction corresponding to a ``solid crystalline state'' would have lower 
energy than that calculated perturbatively beginning with the non-interacting 
electron ``liquid'' ground state. Experimental evidence for this crystalline 
``phase''
was later presented for electrons on the (two dimensional) surface of Helium.
\cite{CCG} Variational calculations by Tanatar and Ceperley \cite{TC} on a 
two dimensional electron system (2DES) supported these observations by 
indicating a phase transition when the electron density $n$ was small: 
$(\pi n)^{-\frac{1}{2}}=37\pm 5$ ( =$r_s$ in units of $\frac{\hbar^2}{me^2}$). 
In recent years, the 2DES has been realized at the interface of 
semiconductor heterostructures. In these experiments, the application of 
a perpendicular magnetic field $B$ provides a second parameter, the filling 
factor $\nu$ ($=\frac{2\pi\hbar c n}{|e|B}$), the variation of which has 
resulted in the experimental observation of a rich phase diagram including 
integer \cite{vK} and fractional quantum Hall states \cite{DT} 
at particular values 
of $\nu$. Using the Lindemann melting criterion, the solid-liquid 
phase boundary (in the $r_s$-$\nu$ plane) for the 2DES has also been computed.
\cite{RP1} This calculation shows that the density at which the liquid to solid 
transition occurs increases monotonically with the magnetic field. It also 
shows that even in the extreme quantum limit ($r_s\rightarrow 0$) the 
solid phase exists below $\nu=0.2$ . 

Several recent experiments have reported on the radiative recombination of 
electrons 
from the 2DES at the junction of a GaAs-AlGaAs single heterostructure with 
acceptor atoms a certain distance from the interface in the GaAs layer. 
\cite{IVK3,IVK1,IVK2,HB}. For certain low values of $\nu$ the photon spectrum 
in these experiments has been interpreted as corresponding to the recombination 
of an electron (of the 2DES) in the Wigner solid phase with localized holes
bound to the acceptor atoms. Accepting this interpretation we calculate here  
the line shape of the spectrum from the Wigner solid and compare it with the 
experimental curve.\cite{IVK1} Our goal is to explicitly investigate whether 
the theoretical consequences of this interpretation agree with the actual 
experimental observations of Ref. \onlinecite{IVK1}. 

The model we adopt is as follows. We consider an electron solid in which an 
unoccupied localized state out of the plane of the 2DES is available for an 
electron to tunnel into. One of the electrons - presumably the one closest to 
it - is assumed to tunnel into the available state, leaving behind a set of 
electrons that is no longer in an equilibrium configuration. This configuration 
may be represented as a distorted solid around some new equilibrium 
configuration, and these distortions may be written as a linear combination 
of the phonon states  of the Wigner crystal. The squared amplitude for each 
final state then represents 
the probability of finding the system in some definite state of the final 
Hamiltonian; the difference in energy between this state and the initial state 
is released as a photon. Thus one expects a broad photon spectrum due to the 
many possible final phonon states.

We consider both the limit of a perfect electron solid for the initial state, 
and one in which several charged impurities may be located in the acceptor 
plane. The motivation for the latter model is that the experiments with which 
we wish to compare - time resolved photoluminescence (PL) spectroscopy - 
measures the 
spectrum only after a waiting time has passed. This not only allows the 
electrons to 
settle down into a quasiequilibrium state, but also allows the acceptors that 
are closest to positions of the electrons in the 2DES to absorb electrons 
from it. Thus, after a sufficient waiting perood, the initially neutral 
acceptor atoms, that are meant to probe the system, become a source of disorder 
for it. The disorder associated with the acceptor positions, therefore, may 
affect the PL spectrum. 

Because the holes closest to electrons are eliminated quickly, we assume that 
the holes in our problem are relatively far from any electrons. In the case of 
a perfect crystalline initial state, we assume that hole to be located near 
an interstitial position of the lattice.\cite{IVK3} For a disordered initial 
state, we search for locations in the plane that are far from any electron 
and presume these are good candidates for locations of holes that may 
contribute to PL long after the initial excitation. In either case, we find 
PL lineshapes  that are in good qualitative agreement with experiment.
\cite{IVK1} However, for a perfect Wigner crystal, we find a PL spectral width 
approximately a factor of 3 narrower than the experimental results, whereas 
our random disorder 
model yields a lineshape a factor of 3 broader than experiment. This 
quantitative disagreement may arise from correlations in the acceptor 
positions which our random disorder model neglects. 

Our approach is unusual in that we allow not just for shifts in the 
electronic positions, but also for changes in the phonon frequencies after the 
recombination event has occurred. This can be a potentially important 
improvement, because the removal of an electron effectively removes a degree 
of freedom, and may introduce localized phonon states in the vicinity of the
(charged) acceptor atom. Previous approaches to sudden switching problems 
such as this have assumed the phonon spectrum to be the same around the initial 
and final equilibrium configurations of the electrons.\cite{JK,Gdm}

This paper is organized as follows. In section II we describe our model in 
details
and explain how the PL spectrum may be derived from it. Section III is a 
discussion of our results, and we conclude with a summary in 
section IV. Because we employ novel theoretical techniques not readily 
available in the literature, our appendices are fairly complete, which, 
however, 
can be skipped unless one is interested in following the technical details 
of our calculation. Detail of the calculation are given in these Appendices.

\section{Theoretical Approach}

The harmonic solid model has been successfully used previously to study 
tunneling between two 2DESs in a strong perpendicular magnetic field. \cite{JK}
The quantitative success of this model in explaining the experimental I-V 
characteristics \cite{EPW} has motivated us to adopt a {\it similar} model 
for computing the PL spectrum. 
We therefore determine the Hamiltonian {\it for the x-y degrees of freedom} 
of the 2DES (assumed to be in the {\it x-y} plane) within the harmonic 
approximation. This approach could be considered to be complimentary to 
the previous Hartree-Fock theory for the phonon shakeup effects on 
PL. \cite{HA}

The system studied consists of a {\it finite} number of electrons in the 
{\it x-y} 
plane within a boundary of the shape of a parallelogram that is commensurate 
with a triangular lattice.  In the model we study there are {\it always} some 
electrons that are pinned - they may be in the {\it x-y} plane and/or below the 
plane corresponding to the charged acceptor atoms.
The electron-electron interactions are however not
purely Coulomb like - this is primarily due to the application of periodic 
boundary conditions which simulates an infinite system by repeating the 
finite system at integral multiples of {\it its} lattice constants. (These 
lattice constants, $\vec a_1$ and $\vec a_2$, correspond to the parallelogram 
that serves 
as a unit cell that contains the entire finite system.) In some of the 
cases studied here the Coulomb interaction has been ``softened'' to 
the form $(r^2+z^2)^{-\frac{1}{2}}$ to account for the finite extent of the 
2DES in the $z$ direction.

The process of electron capture from the
2DES by the acceptor atom is modeled to be a sudden perturbation of the
Hamiltonian corresponding to the {\it x-y} degrees of freedom of the 2DES.
The recombination process is represented through the introduction of a 
fictitious 
parabolic external potential in the final Hamiltonian that confines the 
{\it x-y} 
coordinates of one of the electrons to a position $\vec r_0$
corresponding to the {\it x-y} coordinates of the acceptor atom that is 
capturing the electron. The change in the $z$ coordinate 
of this recombining electron is also reflected as a sudden perturbation  
of the corresponding 
electron-electron interactions. The initial 
(${\cal H}_i$) and final (${\cal H}_f$) Hamiltonians, (corresponding to the 
{\it x-y} degrees of freedom) in the presence of a 
uniform magnetic field in the $z$ direction can therefore be written 
(using the symmetric gauge for the vector potential $\vec A$) as: 

\begin{eqnarray}
{\cal H}_{i(f)}&=&\sum_{k=1}^{N_{i(f)}} \frac{1}{2m^*}  
\left( \vec p_k - \frac{e}{c}
\vec A(\vec r_k) \right)^2 \nonumber \\
&+&\frac{\displaystyle 1}{\displaystyle 2}\sum_{k=1}^{M}\sum_{l=1}^{M}
\tilde V(\vec r_k-\vec r_l,z_{i(f)}^{k,l}) 
+ \lambda_{i(f)}\left(\vec r_c - \vec r_0 \right)^2 \;,  
\end{eqnarray}
\noindent with
\begin{eqnarray}
\vec A(\vec r)&=&
\frac{1}{2}B \hat{e}_z \times \vec r {\mathrm \; \; and, } \nonumber \\
\tilde V(\vec r,z)&=&\frac{e^2}{\epsilon_0} \times \left\{ 
\stackrel {\displaystyle \mathrm{lim.}}{\vec q\rightarrow 0}
\left[ \sum_{\displaystyle \vec R}
\frac{\displaystyle e^{\displaystyle i\vec q\cdot\vec R}}
{\displaystyle |\vec r + z\hat{e}_z + \vec R|} \right. \right. 
\nonumber \\
&&\left. \left. -\frac{ \displaystyle 1} {\displaystyle A} {\displaystyle \int}
\frac{\displaystyle d^2r \, ^\prime e^{\displaystyle i\vec q\cdot
\vec r \, ^\prime}}{\displaystyle |\vec r + z\hat{e}_z +
\vec r \, ^\prime |} \right] 
 -\frac{ \displaystyle \theta(\vec r+z\hat e_z = 0)}
{\displaystyle |\vec r+z\hat{e}_z|} \right\}\;, \nonumber 
\end{eqnarray}
\noindent where
\begin{eqnarray}
\vec R&=&m_1\vec a_1 + m_2 \vec a_2\;,\;\;\;m_{1,2}\in\;\; {\mathrm integers} 
\;\;\;\;{\mathrm and} \nonumber\\
A&=&|\vec a_1 \times \vec a_2| \; .\nonumber
\end{eqnarray} 

\noindent In the above equations, $m^*$ is the effective mass of the electron, 
$\epsilon_0$ 
is the dielectric constant of GaAs. 
$\vec r_{i=c}$ corresponds to the electron (which is necessarily 
one that is {\it not} pinned) that is recombining with a hole at $\vec r_0$. 
The limiting prescription in the expression for $V(\vec r,z)$ is for 
removing the singularity in the lattice sum (the summation over \{$\vec R$\}) 
which is due to the long range nature of the Coulomb interaction. \cite{BM} 
The $\theta$ 
function in this expression is equal to one if the condition in its argument 
is true and zero otherwise - it allows one to represent a particle to be 
interacting with its images in other unit cells, but not with itself. 
$z_{i(f)}^{k,l}$ is the out of plane separation between particles $k$ and $l$ 
in the initial (final) state.
$N_{i(f)}$ is the number of electrons that 
are dynamical degrees of freedom (in the sense that their coordinates can 
``evolve'') and 
$M$ is the total number of electrons including those that are pinned.
Hence, in ${\cal H}_{i(f)}$, in addition to 
\{$z_{i(f)}^{k,l},\; k,l=1\;\mathrm{to}\;M$\}, \{${\vec r_k},\;k=N_{i(f)}+1\; 
\mathrm{to}\;M$\} also serve only as parameters and not dynamical degrees of 
freedom.
The choice of these parameters varies with the different cases studied here
and will be described in detail in the next section. $\lambda_i=0$ and 
$\lambda_f$ is chosen to be a much larger than the ``natural'' scale (of the 
same dimension) in this problem: $e^2n^{1.5}/\epsilon_0$. 
This choice of $\lambda_f$ models the experimental situation in which the 
electron recombines with a screened localized core hole. In principle one would 
like to remove this recombined electron as a degree of freedom in the 
final Hamiltonian, but from a calculational point of view it is easier to 
keep it and introduce the last term in Eq. 1, effectively freezing its 
motion.  The lattice sums in the potential energy 
terms can be evaluated using standard techniques
\cite{RP1,ND,VM} (see Appendix A).  

The harmonic approximation to ${\cal H}_{i(f)}$ (denoted by 
${\cal H}_{i(f)}^h$) is developed by 
expanding the corresponding total potential energy $V$ (=the sum of the last 
two terms in Eq. 1) about the (classical) equilibrium configurations. The 
equilibrium configuration is reached by changing the 
coordinates of all the unpinned electrons such that the forces on them 
become zero. The coordinates of electrons chosen at the beginning of this 
``evolution'' vary with the different cases studied here and will be 
described in detail in the next section. The algorithm for this evolution is 
due to Schweigert 
and Peeters.\cite{SP} This algorithm updates the coordinates by finding the 
minimum of $V$ when expressed to second order in the 
coordinate increments {\it i.e.} if $V$ is expressed as a function 
of the coordinates of all the unpinned electrons ($q_i$,$1\leq i\leq 2N$,  
$N=N_i$ or $N_f$) , 
then it may be written approximately ($V^a$), to second order in the 
coordinate increments from the current position ($q_i^t$) as:

\begin{eqnarray}
V^a({\mathbf q})&=&V({\mathbf q}^t) - ({\mathbf f}^t)^T
({\mathbf q}-{\mathbf q}^t) 
+ \frac{1}{2}({\mathbf q}-{\mathbf q}^t)^T {\mathbf D}^t 
({\mathbf q}-{\mathbf q}^t)\;, \nonumber \\ 
{\mathrm \; \; with} &&\nonumber \\
({\mathbf f}^t)^T&=&- 
    \left[ \frac{ \displaystyle \partial V({\mathbf q}) }
                { \displaystyle \partial {\mathbf q}    } \right]_
{ {\mathbf q} = {\mathbf q}^t} 
{\mathrm \; \; and \; \;} 
{\mathbf D}^t = 
\left[\frac{ \displaystyle \partial^2 V({\mathbf q}) }
       {\displaystyle \partial {\mathbf q} \partial {\mathbf q}^T} \right]_
{ {\mathbf q} = {\mathbf q}^t} \;,\nonumber
\end{eqnarray}

\noindent where ${\mathbf q}$ is a column vector whose components are the 
coordinates $q_i$ and therefore ${\mathbf f}$ becomes a column vector 
whose components are the forces and ${\mathbf D}$ becomes the (symmetric) 
dynamical matrix. 
The lattice sums appearing in ${\mathbf f}$ and ${\mathbf D}$ are again 
evaluated using standard techniques \cite{RP1,ND,VM} (see Appendix A). 
The updated coordinate (${\mathbf q}^{t+1}$) is determined by minimizing 
$V^a({\mathbf q})$. However, far away from the equilibrium configuration 
the matrix ${\mathbf D}^t$ may not be positive definite and hence 
$V^a({\mathbf q})$ may not have a minimum. Therefore the matrix 
${\mathbf D}^t$ is (arbitrarily) changed to ${\mathbf D^\prime}^t = 
{\mathbf D}^t + \eta^t{\mathbf I}$ where $\eta^t$ is a convergence parameter 
suitably chosen such that ${\mathbf D^\prime}^t$ is positive definite. We have 
chosen it to be proportional to the magnitude of the largest component in 
${\mathbf f}^t$. In most cases this modification of ${\mathbf D}$ resulted 
in a stable evolution. However in some cases it failed {\it i.e.} 
${\mathbf D^\prime}^t$ was nearly singular leading to numerical instabilities. 
In such cases, ${\mathbf D^\prime}^t$ was more carefully constructed after 
diagonalizing ${\mathbf D}^t$: if ${\mathbf D}^t={\mathbf M}\Lambda
{\mathbf M^T}$ (where $\Lambda$ is the diagonal matrix consisting of the 
eigenvalues of ${\mathbf D}^t$) then ${\mathbf D^\prime}^t={\mathbf M}
f(\Lambda){\mathbf M^T}$ such that $f(\Lambda)$ is a diagonal 
positive definite matrix with the function $f$ chosen to satisfy $f(\Lambda) 
\rightarrow \Lambda$ as convergence is reached. Hence the algorithm for 
updating the coordinates becomes:  

\begin{eqnarray}
{\mathbf q}^{t+1} = {\mathbf q}^{t} + \left( {\mathbf D^\prime}^t \right)^
{-1} {\mathbf f}^t \;. \nonumber
\end{eqnarray}

\noindent The above algorithm is iterated until convergence is reached - the 
root mean squared force per electron becoming smaller than a chosen value. 
This yields the equilibrium configuration ${\mathbf q}^{eq}$.

The process of determining the equilibrium configuration provides all 
the necessary parameters (from the potential energy terms) for the harmonic 
approximation to the Hamiltonian. 
In this approximation the potential $V$ is replaced by its approximate form 
$V^a({\mathbf q})$, with the expansion of $V^a({\mathbf q})$ around the 
equilibrium configuration ${\mathbf q}^{eq}$. Hence, since ${\mathbf f^{eq}}=0$,
the parameters that 
enter the approximate Hamiltonian are contained in 
the dynamical matrix ${\mathbf D^{eq}}$. This matrix is positive definite since 
the equilibrium configuration reached is {\it stable}. There are no modes 
corresponding to neutral equilibrium since global translational
invariance is broken by the pinned electrons. There is no global rotational 
invariance even in the absence of pinning due to the chosen periodic boundary 
conditions. 
Further, due to the linear variation 
of the vector potential $A(\vec r)$ with $\vec r$, the approximated 
initial (${\cal H}^h_i$) and final Hamiltonians (${\cal H}^h_f$) can 
be written (with $\epsilon^m_{i(f)}=V({\mathbf q}_{i(f)}^{eq})$) as:

\begin{eqnarray}
{\cal H}^h_{i(f)}& =& \frac{1}{2m^*} \left( {\mathbf \Pi - Bq} \right)^T
\left( {\mathbf \Pi - Bq} \right)  \nonumber \\
&& + \frac{1}{2} \left( {\mathbf q - q}_{i(f)}^{eq} \right)^T 
{\mathbf D}_{i(f)}^{eq} \left( {\mathbf q - q}_{i(f)}^{eq} \right)+
\epsilon^m_{i(f)} \;,
\end{eqnarray} 

\noindent where ${\mathbf \Pi}$ is a column vector whose components 
$\Pi_k$ are the  (canonical) momenta corresponding respectively 
to the coordinates $q_k$ ($1\leq k \leq 2N_f$) and ${\mathbf B}$ is 
a (antisymmetric) matrix corresponding to the vector potential. Note that 
the harmonically approximated Hamiltonians have an equal number of dynamical 
degrees of freedom ($2N_f$). 
For the disordered systems investigated, in some cases we chose
to randomly pin a small number of electrons after the initial
equilibrium configuration had been found.  This is meant to
roughly model the effect of pinned in-plane charged impurities \cite{RMS}
that are not due to the charged acceptors.  In such cases $N_f< N_i$.

Gauge freedom 
may now be exploited to change $\vec A(\vec r_k)$ to the form 
$\vec A(\vec r_k) = \frac{1}{2}B \hat{e}_z \times \left( \vec r_k - (\vec r_k)^
{eq}_i \right)$ where $(\vec r_k)^{eq}_i$ corresponds to the equilibrium 
configuration 
${\mathbf q}^{eq}_i$ of the initial Hamiltonian. Further defining 
${\mathbf \Phi}_i = {\mathbf q - q}^{eq}_i$, 
${\mathbf \Phi}_f = {\mathbf q - q}^{eq}_f$, and the corresponding canonical 
momenta ${\mathbf \Pi}_i = {\mathbf \Pi}$, ${\mathbf \Pi}_f = {\mathbf \Pi - B}
\left({\mathbf q}^{eq}_i - {\mathbf q}^{eq}_f \right)$ , ${\cal H}^h_i$ and 
${\cal H}^h_f$  may be written as: 

\begin{eqnarray}
\lefteqn{  {\cal H}_{i(f)}^h= \epsilon^m_{i(f)} + \frac{1}{2m^*} \times}
\nonumber \\
&\begin{array}{ccc}
  \left[{\mathbf \Phi}_{i(f)}^T {\mathbf \Pi}_{i(f)}^T \right] &
  \left[ \begin{array}{cc}
        m^*{\mathbf D}_{i(f)}^{eq}+{\mathbf B}^T{\mathbf B} & -{\mathbf B}^T \\
           -{\mathbf B}                            & {\mathbf I} 
         \end{array}
  \right] & 
  \left[ \begin{array}{c}
           {\mathbf \Phi}_{i(f)} \\
           {\mathbf \Pi}_{i(f)} 
         \end{array}
  \right]
                \end{array} \;,  \nonumber
\end{eqnarray}
\noindent  with 
\begin{eqnarray}
\left[ 
       \begin{array}{c}
         {\mathbf \Phi}_{f} \\
         {\mathbf \Pi}_{f} 
       \end{array}
\right] 
        = 
\left[
       \begin{array}{c}
         {\mathbf \Phi}_{i} \\
         {\mathbf \Pi}_{i}
       \end{array}
\right] 
    + 
\begin{array}{cc}
  \left[ \begin{array}{cc}
           {\mathbf I} & -{\mathbf I} \\
           {\mathbf B} & -{\mathbf B}
          \end{array}
  \right] &
  \left[ \begin{array}{c}
           {\mathbf q}_i^{eq}  \\
           {\mathbf q}_f^{eq}
          \end{array}
  \right] 
\end{array} \;.  
\end{eqnarray}

The harmonic Hamiltonians thus developed can be considered to be functions 
of classical variables or the corresponding quantum mechanical operators 
(with electrons being spinless). 
In either case, by a linear canonical transformation of the phase space 
variables , they can be written as a sum of uncoupled (normal) modes (see 
Appendix B). Hence  

\begin{eqnarray}
{\cal H}_{i(f)}^h&=&\epsilon^m_{i(f)} + \\   
&\frac{\displaystyle 1}{\displaystyle 2}&
\begin{array}{ccc}
  \left[{\mathbf \Psi}_{i(f)}^T {\mathbf \Xi}_{i(f)}^T \right] &
  \left[ \begin{array}{ll}
          {\mathbf \Omega}_{i(f)} & {\mathbf 0} \\
          {\mathbf 0}            & {\mathbf \Omega}_{i(f)}  
         \end{array}
  \right] & 
  \left[ \begin{array}{c}
           {\mathbf \Psi}_{i(f)} \\
           {\mathbf \Xi}_{i(f)} 
         \end{array}
  \right]
                \end{array}   \;,  \nonumber \\
{\mathrm with} &&
\left[ 
       \begin{array}{c}
         {\mathbf \Psi}_{i(f)} \\
         {\mathbf \Xi}_{i(f)} 
       \end{array}
\right] 
       =
\left[
        {\mathbf C}_{i(f)}
\right] 
  \left[ \begin{array}{c}
           {\mathbf \Phi}_{i(f)}  \\
           {\mathbf \Pi}_{i(f)}
          \end{array}
  \right]  \; . 
\end{eqnarray}

\noindent Since the transformation ${\mathbf C}_{i}$ (${\mathbf C}_{f}$) 
is canonical, the components of ${\mathbf \Xi}_{i}$ (${\mathbf \Xi}_{f}$) are 
the conjugate momenta corresponding to the components of the 
${\mathbf \Psi}_{i}$ (${\mathbf \Psi}_{f}$) with ${\mathbf \Omega}_{i}$ 
(${\mathbf \Omega}_{f}$) being a diagonal matrix consisting of the normal 
mode frequencies of ${\cal H}_i^h$ (${\cal H}_f^h$).

The Planck's constant $\hbar$ may now be explicitly introduced into the 
Hamiltonians ${\cal H}_{i(f)}$ by assuming that ${\mathbf \Xi}_{i(f)}$ and 
${\mathbf \Psi}_{i(f)}$ are quantum mechanical operators. Hence, defining the 
column vectors consisting of lowering (${\mathbf a}_{i(f)}$) and raising 
(${\mathbf a}_{i(f)}^\dagger $) operators: 

\begin{eqnarray}
\left[
       \begin{array}{c}
         {\mathbf a}_{i(f)} \\
         {\mathbf  a}_{i(f)}^\dagger
       \end{array}
\right]
        =
\begin{array}{cc}
   \left( \frac{\displaystyle 1}{\displaystyle 2 \hbar} \right)^\frac{1}{2}
   \left[
          \begin{array}{cr}
            {\mathbf I} & i{\mathbf I} \\
            {\mathbf I} & -i{\mathbf I}
          \end{array}
   \right] &
   \left[
          \begin{array}{c}
             {\mathbf \Psi}_{i(f)} \\
             {\mathbf \Xi}_{i(f)}
          \end{array}
   \right]
\end{array} \;, 
\end{eqnarray}

\noindent the Hamiltonians may be written as 

\begin{eqnarray}
{\cal H}_{i(f)}^h&=& \epsilon^m_{i(f)} + \\
&&\begin{array}{ccc}
  \frac{\displaystyle \hbar}{\displaystyle 2}
  \left[\tilde{\mathbf a}_{i(f)} \tilde{\mathbf a}_{i(f)}^\dagger \right] &
  \left[ \begin{array}{ll}
          {\mathbf 0}     & {\mathbf \Omega}_{i(f)}  \\
          {\mathbf \Omega}_{i(f)} & {\mathbf 0}
         \end{array}
  \right] &
  \left[ \begin{array}{c}
           {\mathbf a}_{i(f)} \\
           {\mathbf a}_{i(f)}^\dagger
         \end{array}
  \right]
                \end{array} \;, \nonumber 
\end{eqnarray}
\noindent with (using Eqs. 3,5 and 6)

\begin{eqnarray}
\left[
       \begin{array}{c}
         {\mathbf a}_{f} \\
         {\mathbf a}_{f}^\dagger
       \end{array}
\right] = {\mathbf T}  
  \left[ \begin{array}{c}
           {\mathbf a}_i  \\
           {\mathbf a}_i^\dagger
          \end{array}
  \right] + {\mathbf w} \;,
\end{eqnarray}
\noindent where ${\mathbf T}$ is a matrix of the form
\begin{eqnarray}
{\mathbf T} = 
\left[ \begin{array}{ll}
           {\mathbf U} & {\mathbf V} \\
           {\mathbf V}^* & {\mathbf U}^*
          \end{array}
  \right]
   =
   \frac{\displaystyle 1}{\displaystyle 2 }
   \left[
          \begin{array}{cr}
            {\mathbf I} & i{\mathbf I} \\
            {\mathbf I} & -i{\mathbf I}
          \end{array}
   \right] 
   {\mathbf C}_f {\mathbf C}_i^{-1}
   \left[
          \begin{array}{rr}
            {\mathbf I} & {\mathbf I} \\
            -i{\mathbf I} & i{\mathbf I}
          \end{array}
   \right]                    \;,                  
\end{eqnarray}
\noindent and ${\mathbf w}$ is a vector of the form
\begin{eqnarray}
{\mathbf w} &=& 
\left[
       \begin{array}{l}
         {\mathbf \Delta} \\
         {\mathbf \Delta}^*
       \end{array}
\right] \\
 &=&
   \left( \frac{\displaystyle 1}{\displaystyle 2 \hbar} \right)^\frac{1}{2}
   \left[
          \begin{array}{cr}
            {\mathbf I} & i{\mathbf I} \\
            {\mathbf I} & -i{\mathbf I}
          \end{array}
   \right]
  {\mathbf C}_f
   \left[
          \begin{array}{cc}
            {\mathbf I} & -{\mathbf I} \\
            {\mathbf B} & -{\mathbf B}
          \end{array}
   \right]  
   \left[ \begin{array}{c}
           {\mathbf q}_i^{eq}  \\
           {\mathbf q}_f^{eq}
          \end{array}
   \right] \;. \nonumber
\end{eqnarray}

We now proceed to describe the theory of PL. The initial states 
$|\psi_i\rangle$ are assumed to be eigenstates of ${\cal H}_i^h$ 
(with ${\cal H}_i^h |\psi_i\rangle = E_i|\psi_i\rangle$) whose distribution 
is determined by the temperature $\beta= (k_bT)^{-1}$. After the sudden 
perturbation, in which the Hamiltonian changes from 
${\cal H}_i^h$ to ${\cal H}_f^h$, the initial state is assumed to collapse into 
a final state that is an eigenstate $|\psi_f\rangle$ of ${\cal H}_f^h$ (with 
${\cal H}_f^h|\psi_f\rangle = E_f|\psi_f\rangle$) with a probability  
$|\langle\psi_f|\psi_i\rangle|^2$. In this process a
photon of frequency $\omega$ is emitted. Assuming conservation of energy 
across the sudden perturbation, $\hbar\omega = E_i - E_f$. Hence the 
probability density of the photon frequency 
${\cal P}(\omega)$ is given by:
\begin{eqnarray}
{\cal P}(\omega)&=&\left[ \sum_i e^{-\beta E_i} \right]^{\displaystyle -1}
\times  \\
&&{\displaystyle \sum_i} e^{-\beta E_i} {\displaystyle \sum_f} 
|\langle\psi_f|\psi_i\rangle|^2 \delta
\left( \hbar^{-1}(E_i-E_f) - \omega \right) \;. \nonumber 
\end{eqnarray}
\noindent ${\cal P}(\omega)$ may be calculated from its Fourier transform 
$\tilde{\cal P}(t)$:
\begin{eqnarray}
{\cal P}(\omega) = \frac{\displaystyle 1}{\displaystyle 2\pi} 
{\displaystyle \int}_{-\infty}^{+\infty} dt e^{\displaystyle -i\omega t} 
\tilde{\cal P}(t) \;,
\end{eqnarray}
\noindent with  $\tilde{\cal P}(t)$ given by (inverting Eq. 12 and using 
Eq. 11):
\begin{eqnarray}
\tilde{\cal P}(t)=\frac{\displaystyle 
\sum_i e^{-\beta E_i} \sum_f|\langle\psi_f|\psi_i\rangle|^2 
e^{i\hbar^{-1}(E_i-E_f)t}} {\displaystyle \sum_i e^{-\beta E_i} }  \;.
\end{eqnarray}
\noindent Rewriting Eq. 13 using the Hamiltonians ${\cal H}_{i(f)}^h$:
\begin{eqnarray}
\tilde{\cal P}(t)&=&\left[ \sum_i \langle\psi_i|e^{\displaystyle -\beta 
{\cal H}_i}|\psi_i\rangle \right]^{\displaystyle -1} \times \\
\sum_i \langle\psi_i| \sum_f &&|\psi_f\rangle \langle\psi_f|
e^{\displaystyle -i\hbar^{-1} {\cal H}_f^ht} 
e^{\displaystyle i\hbar^{-1} {\cal H}_i^h(t+i\hbar\beta)}
|\psi_i\rangle \;. \nonumber 
\end{eqnarray}
\noindent Since $|\psi_i\rangle$ and $|\psi_f\rangle$ constitute an 
orthonormal bases, Eq. 14 may be written in terms of trace ($Tr$) operations 
that are basis independent: 
\begin{equation}
\tilde{\cal P}(t) = 
\frac{\displaystyle
Tr \left[
e^{\displaystyle -i\hbar^{-1} {\cal H}_f^ht}
e^{\displaystyle i\hbar^{-1} {\cal H}_i^h(t+i\hbar\beta)}
   \right] }
{\displaystyle
Tr \left[ e^{\displaystyle -\beta {\cal H}_i}   \right]  } \;.
\end{equation}
\noindent ${\cal P}(\omega)$ is then computed using Eq. 12 . 
In order to compare ${\cal P}(\omega)$ to the experimental line shape of 
the intensity $I(\omega)$ we make the final assumption that:
\begin{equation}  
I(\omega) \propto {\cal P}(\omega) \;.
\end{equation}

The above relationship (after substituting for ${\cal P}(\omega)$ 
using Eq. 11) may be derived using first order 
time dependent 
perturbation theory with a time independent Hamiltonian that includes the 
photon degrees of freedom together with the $z$ degree of freedom for the 
recombining electron (see Appendix C). Finally, we note that our model for 
the final state of the recombined electron formally retains it as a degree 
of freedom, albeit with a very large spring constant $\lambda_f$. In practice
this means that the spectrum ${\cal P}(\omega)$ consists of a series of 
highly separated peaks, each corresponding to a different state of the 
recombined electron plus phonons in the remaining lattice. Physically, only 
the state in which the electron is most strongly localized at the location of 
the core hole - {\it i.e.}, the lowest energy state of the harmonic potential
due to the $\lambda_f$ in Eq. 1 - is truly relevant. Because of the separation 
of energy scales between the ``$\lambda$ modes'' and the lattice phonon 
modes, one may easily identify the highest energy peak in ${\cal P}(\omega)$ 
as the experimentally relevant spectrum. From a computational point of view, 
one would like to eliminate the other peaks, as they are unphysical and can 
consume much cpu time in their computation. This can be accomplished by 
introducing an imaginary component {\it i.e.}, a broadening to the normal mode 
frequencies. The 
precise way in which this is done, along with several other practical issues 
in our computation and approximation scheme, is discussed 
in Appendix D. Details of the computation of the traces
 appearing in Eq. 15 are given in Appendix E.

\section{Results and Discussion}

We now present out theoretical results corresponding to the experiment of 
Kukushkin {\it et al} \cite{IVK1} who identify a particular ``late time'' 
spectrum in their time 
resolved PL spectra as corresponding to the recombination of 
electrons from a Wigner solid with holes bound to acceptor atoms a certain 
distance away from the 2DES. For comparing our calculations with their 
observations, we set the electron density $n$ to 
$5.3\times10^{10}/$cm$^2$, the filling factor $\nu$ to $0.1337$ 
(a magnetic field $B\approx16.4$ Tesla), the temperature $T$ to 
$45$ mK and the distance between the 2DES and the acceptor atoms ($z_0$)
 to $300 \AA$. The effective mass 
parameter $m^*$ is set to $0.068m_e$ and the dielectric constant $\epsilon_0$ 
is set to $12.8$ which correspond to the AlGaAS-GaAs heterostructure. We 
adopt natural units defined by $n$, $e^2/\epsilon_0$ and $m^*$. Therefore the 
natural length scale is  
($=1/\sqrt{n}$) $\approx 434\;\AA$, the 
natural frequency scale is ($=\sqrt{e^2n^{1.5}/\epsilon_0m^*}$) 
$\approx 1.88$ THz and the 
natural energy scale is ($=e^2n^{0.5}/\epsilon_0$) $\approx 2.59$ meV. 
In these units, the cyclotron frequency $\approx 22.5$, $\hbar$  
$\approx 0.479$ and the inverse temperature ($\beta$) $\approx 668$.

It must be noted that the experimental late time spectral ``width'' \cite{IVK1}
(the range of photon {\it frequencies} for which the intensity can be 
distinguished from being $\approx 0$) is $\sim 3.0$ in the natural units 
chosen. We study four 
cases here in an attempt to reproduce this width and thereby identify the
experimentally relevant one. In one of these cases we also attempt to 
give an alternative interpretation to the experimental observation \cite{IVK1} 
of a ``double peak'' during continuous illumination. As the experimental 
spectra have the intensity plotted against increasing photon wavelength,
\cite{IVK1} we 
show the calculated spectra with decreasing photon frequency. The comparison 
of line shape can then be directly made since in the experiment the spread in 
photon frequency is a very small fraction ($\sim 2\times10^{-3}$) of the 
average frequency. As mentioned previously, the binding energy of the 
electron to the hole, which determines the position of the spectrum along the 
frequency axis, cannot be determined in our model and has been set to  
zero when presenting the results. Therefore the calculated peak always 
appears in the negative frequency domain.

\subsection{Perfect Wigner lattice with Coulomb interactions}

In this case the initial (classical) equilibrium configuration corresponds to 
a perfect 
triangular lattice. Because of the translational invariance of a perfect, 
unpinned Wigner crystal, one cannot apply the harmonic approximation 
consistently to this case: the sudden change in potential will introduce 
a large coherent motion of the center of mass in the subsequent dynamics of the 
system. This behavior is unphysical, because in practice a Wigner 
crystal will always be pinned by disorder. We therefore assume that the 
electrons at the boundary of our supercell are actually pinned, and are 
not dynamical degrees of freedom. This allows the harmonic approximation 
to be applied consistently. We will show below that the PL spectrum converges 
rapidly as the system size increases, so that the pinning at the boundary 
does not affect our final results. 

\begin{figure}
 \vbox to 10cm {\vss\hbox to 7cm
 {\hss\
   {\includegraphics{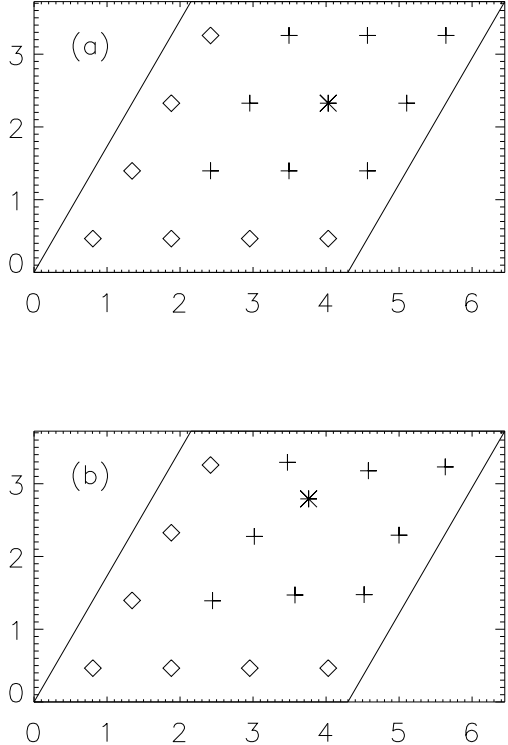}
   }
  \hss}
 }
\caption{Equilibrium {\it x-y} configuration of the perfect Wigner lattice (a)
corresponding to the initial Hamiltonian ${\cal H}_i$ (Eq. 1) and the same
of the
perturbed lattice (b) corresponding to the final Hamiltonian ${\cal H}_f$.
Pinned and unpinned electrons are shown by $\diamond$ and $+$ respectively.
The figures correspond to the case when the number of unpinned electrons
$N_i=N_f=9$ with the total number of electrons $M=16$.
The ``central'' (unpinned) electron is shown by $\ast $.
The axes are labeled in units of the natural length scale of the problem:
(total electron density)$^{-1/2}$.}
\end{figure}

Here, all the 
electrons interact through the ``ideal'' Coulomb interaction. The initial 
configuration can therefore be directly constructed by laying down the 
electrons at 
lattice positions or arrived at by evolving an initial 
configuration in which only the pinned electrons are laid down at lattice 
positions with the rest of the electrons being randomly placed. The final 
configuration, obtained through evolution beginning with the perfect Wigner 
lattice,  corresponds to confining the ``central'' electron (see Fig. 1) 
to {\it one particular} point on the boundary of its Wigner-Seitz cell 
(corresponding to the  prefect lattice).  This choice is motivated 
by the  interpretation\cite{IVK1} that the late time 
PL spectrum corresponds to recombination events in which the 
distance between the hole (bound to the acceptor) and the recombining electron 
is maximal. It must be noted that to determine the final equilibrium 
configuration it is not appropriate to begin with a random configuration of 
unpinned electrons since then the equilibrium configuration reached may 
involve exchanges of electrons with respect to the perfect lattice. 
In this work, we assume that the initial state [Fig. 1(a)] may be regarded 
as a deformation of a classical equilibrium that is closest in configuration 
to this state [Fig. 1(b)], and that the subsequent motion of the electrons 
is due to the vibrations around the latter state. While final states in which 
electrons are exchanged are in principle relevant, such exchanges in practice 
contribute little to the PL spectrum. \cite{MZ}

\begin{figure}
 \vbox to 10.25cm {\vss\hbox to 7cm
 {\hss\
   {\includegraphics{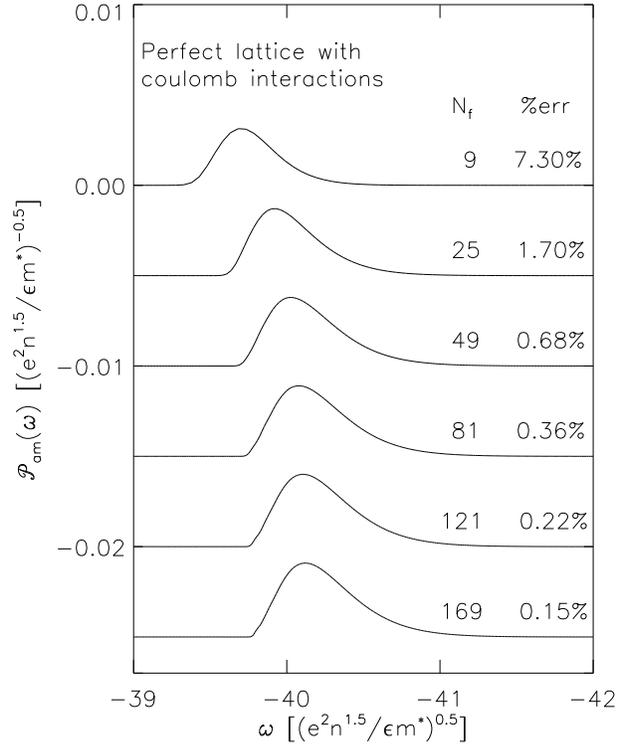}
   }
  \hss}
 }
\caption{ Calculated spectra from transitions similar to that in
Fig. 1 (which corresponds to the number of unpinned electrons ($N_f$)
being equal to nine). For each value of $N_f$, flat regions correspond to 
the zero of ${\cal P}_{am}(\omega)$. $\%err$ is
the increase in calculated spectral width due to approximations made to
get a continuous curve for the finite $N_f$. The spectral width saturates
(around $N_f=25$) to $\sim \frac{\displaystyle 1}{\displaystyle 3}$ of that
observed by Kukushkin {\it et al}.$^2$  However, a qualitative feature
of the
experimental spectrum is present in the theoretical results: a faster rising 
edge as compared to the falling edge.  }
\end{figure}

We can therefore fix the parameters in ${\cal H}_{i(f)}$ (Eq. 1):
$N_i=N_f=P^2$, $M=(P+1)^2$ (Fig. 1 corresponds to the case when $P=3$),
$z_i^{k,l}=0$ (corresponding to all electrons initially being in the {\it x-y} 
plane) and $z_f^{k,l}=z_0(\delta_{l,c}-\delta_{k,c})$ (corresponding to all the 
electrons finally being in the {\it x-y} plane except for the central electron 
$c$ being below the plane at the position of the acceptor atoms). $\vec r_0$ 
and other parameters 
corresponding to the {\it x-y} position of the pinned electrons ($\vec r_i,\;
\;i=N_{i(f)}+1\;\;{\mathrm to}\;\; M$) can be inferred from Fig. 1. 
We have studied the cases with 
$P=[3,5,7,9,11,13]$. In each of these cases we have chosen the parameter 
$\Gamma_0$ (a broadening parameter introduced to make the phonon density of 
states continuous; see Appendix D) to be equal to the smallest phonon 
frequency in ${\cal H}_i^h$. 
These values (in units of $\sqrt{e^2n^{1.5}/\epsilon_0m^*}$) 
are $[7.59,4.49,3.08,2.30,1.82,1.49]\times
10^{-2}$ (corresponding to the cases with $P=[3,5,7,9,11,13]$ respectively).
It must be noted that all the electrons 
including the pinned ones are included in the computation of the density $n$.

Fig. 2 shows the calculated spectra. The error in the spectral width is
due to the approximation of replacing the delta function in Eq.11 by a
Gaussian of width $\Gamma_0$.
Note that the spectral width seems to have saturated beyond $N_f\approx 25$ 
but it is only $\sim 1/3$
of that experimentally observed.\cite{IVK1} A qualitative feature of
the experimental spectrum is however reproduced: a faster rising edge as  
compared to the falling edge.

\subsection{Perfect Wigner lattice with softened Coulomb interactions}

We now soften all the in-plane electron interactions in an attempt to 
see if it results in a broader spectrum than the previous case. The 
softened interaction is of the form $1/\sqrt{r^2 + \bar{z}_0^2}$. 
$\bar{z}_0$ is chosen to be $150\AA$ - a significant fraction of the 
inter-electron distance. This form of the interactions is motivated 
by the fact that the electrons are not strictly confined to 
two dimensions at the interface of the GaAs-AlGaAs heterostructure, there is a 
finite extent in the $z$ direction of the wavefunction confining the 2DES. 
This particularly simple form of the Coulomb interaction incorporating the 
finite width of the electron layer has been quantitatively successful in 
obtaining the fractional quantum Hall excitation gap energies in the 
2DES, \cite{ZD} which motivates us to apply this softened interaction in 
the Wigner crystal PL calculation. 
The procedure to obtain the initial and final equilibrium configurations  
remains the same as before. These configurations are similar to that shown 
in Fig. 1 - there being a ``wall'' of pinned  electrons on the boundary. Only 
one system size is studied with 
$N_{i(f)}=7^2$, $M=8^2$, $z_i^{k,l}=\bar{z}_0(1-\delta_{k,l})$ 
and $z_f^{k,l}=\bar{z}_0(1-\delta_{k,l}) + (z_0 - \bar{z}_0)
|\delta_{k,c}-\delta_{l,c}|$ where $c$ again corresponds to the central 
electron. The parameters $\vec r_0$ and \{ $\vec r_i,\;\;i=N_{i(f)}+1\;\;
{\mathrm to}\;\; M$\} can be inferred from Fig. 1. 
Again, $\Gamma_0$ has been 
chosen to be the smallest phonon frequency 
of ${\cal H}_i^h$ ($=2.52\times 10^{-2} 
\sqrt{e^2n^{1.5}/\epsilon_0m^*}$).

Fig. 3 shows the calculated spectrum. The width is 
slightly reduced (still roughly $1/3$ of the
experimental value) as compared to the case with ideal Coulomb interactions.
This can be attributed to a faster falling edge making it less 
similar to the experimental spectrum.\cite{IVK1}
We conclude that only softening the Coulomb interaction by the finite layer 
thickness of the 2DES cannot account for the experimental spectral width. 
This is understandable, because the spectral width of the PL spectrum is 
primarily determined by the physics of the recombination process at the 
acceptor sites and not by the details of the electron-electron interaction.

\begin{figure}
 \vbox to 8.4cm {\vss\hbox to 7cm
 {\hss\
   {\includegraphics{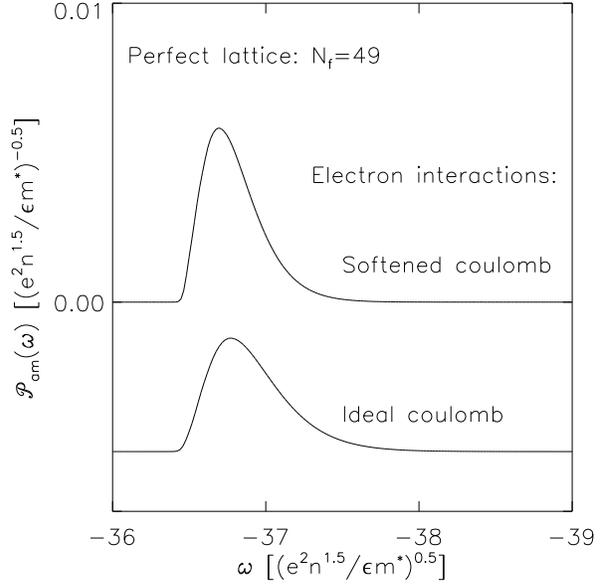}
   }
  \hss}
 }
\caption{ Calculated spectrum from transition similar to that in
Fig. 1 (which corresponds to the number of unpinned electrons $N_f$
being equal to nine) with softened Coulomb interaction. For 
comparison the spectrum corresponding to ideal Coulomb electron interaction 
is included. It has been shifted to the left to make the photon energy 
corresponding to the 
difference in the ground state energies coinside with the same point of the 
first curve. Softening the electron interactions results in the spectrum being
less similar to the experimental curve: the width is slightly reduced 
( continues to be $\sim 1/3$ of the 
experimental value.) and it has a faster falling edge.  }
\end{figure}

\subsection{Perfect Wigner lattice with recombination position averaging}

Since the ``softened'' Coulomb interaction does not lead to any improvement 
in the agreement between theory and experiment, we now revert back to 
``ideal'' Coulomb interaction for in-plane electrons 
and consider averaging spectra corresponding to different {\it x-y} positions 
of the recombination center ($\equiv$ acceptor atom position) relative 
to the initial position of the recombining electron. The 
averaging is carried out assuming that the {\it x-y} distribution of acceptor 
atoms is uniform. This results in a uniform distribution for the position 
of the recombination center within the Wigner-Seitz cell of the recombining 
electron. It models the situation for which the PL is 
not time resolved,\cite
{IVK1,HB} so that no particular final position of the recombining electron 
is favored.
Symmetries can be exploited to reduce the actual region of 
the Wigner-Seitz cell that needs to be explored. For configurations of the 
type shown in Fig. 1 (a), due to the presence of the ``wall'' of pinned 
electrons on the boundary, there are 

\clearpage

\noindent only two reflection symmetries (each along 
a diagonal of the system) for the recombination positions corresponding 
to the central electron. Therefore, to get the average spectrum, 
$1/4$ the area of the Wigner-Seitz cell 
needs to be explored. However, if there was a hexagonal boundary of 
pinned electrons around the recombining electron then there would be 
six-fold together with reflection symmetries along the diagonals 
which would reduce the region 
needed to be explored to
$1/12$ the area of the Wigner-Seitz cell.
We therefore consider such a configuration (Fig. 4).

\widetext

\begin{figure}
 \vbox to 10.5cm {\vss\hbox to 18cm
 {\hss\
   {\includegraphics{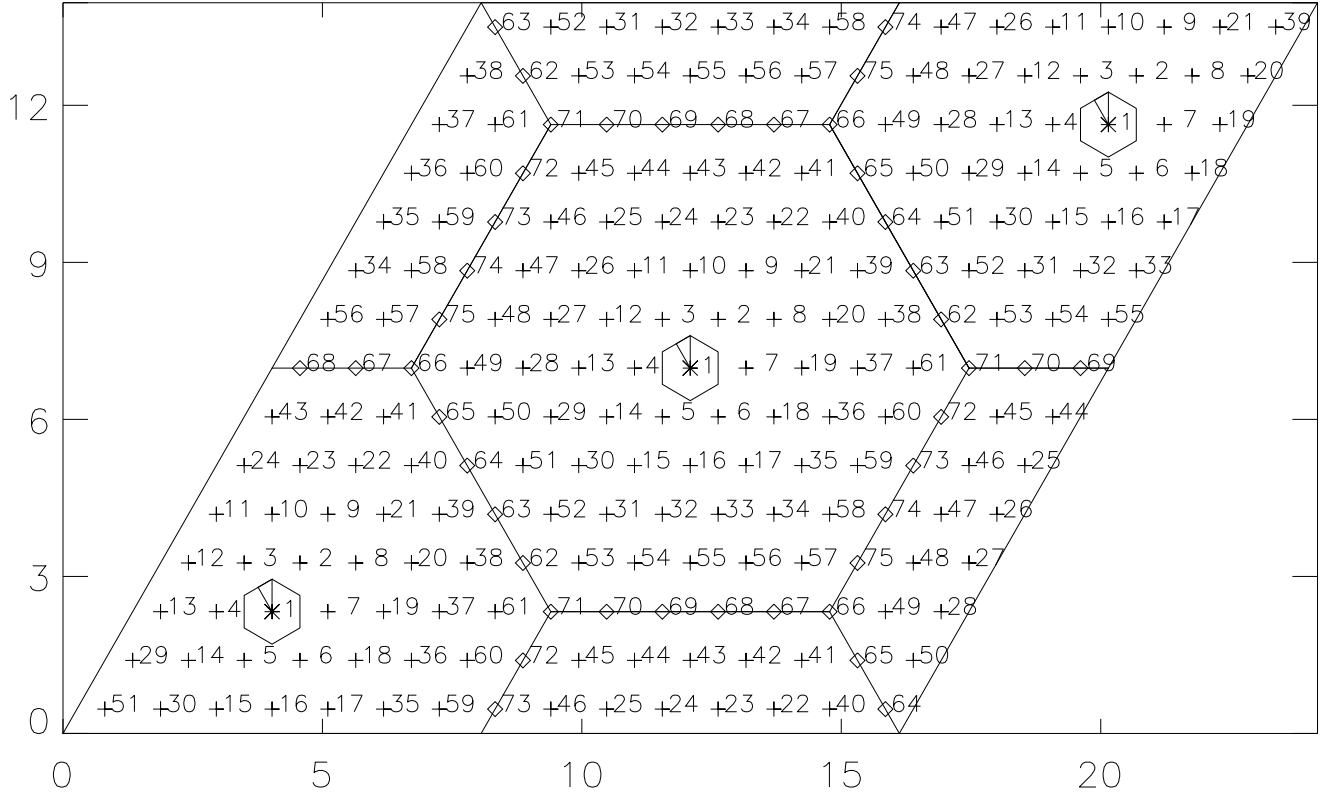}
   }
  \hss}
 }
\caption{ The figure shows the initial equilibrium configuration of 
the system with a hexagonal wall of pinned electrons. Electrons with the  
same number are considered equivalent. This, in addition to application of 
periodic boundary conditions about the parallelogram, results in the hexagonal 
system being repeated periodically - three copies of which are embedded in the 
parallelogram. Therefore the Wigner-Seitz cell of the 
``central'' recombining electron (numbered 1) has all the symmetries of a 
triangular lattice. This reduces the region of different recombination 
positions to be explored to $\frac{\displaystyle 1}{\displaystyle 12}$ the 
area of the Wigner-Seitz cell - the triangle within the hexagon around the 
electron numbered 1. 
 }
\end{figure}

\narrowtext

Fig. 4 shows the construction of the system with a hexagonal wall of pinned 
electrons. Three such copies are needed for embedding into 
a parallelogram - then the usual periodic boundary conditions about the 
parallelogram results in the periodic repetition of the hexagonal system. 
The initial equilibrium configuration is directly constructed as it corresponds
 to a perfect triangular lattice. The final equilibrium configuration is 
arrived at through evolution beginning with this configuration. However, as 
electrons with the same number (see Fig. 4) are considered equivalent, the 
corresponding ``self interaction'' terms are neglected during the computation 
of the forces and the dynamical matrix.  Nevertheless, they are included in the 
computation of the total energy - one third of which is the actual energy of 
the hexagonal system. During evolution, the forces and the dynamical matrix 
need to be computed corresponding to only the unpinned electrons within 
{\it any one} hexagon (interactions between all inequivalent electrons
 are taken into account). By symmetry they are the same for the corresponding 
unpinned electrons in the other two hexagons. The displacements are therefore  
computed for only one set of unpinned electrons but it is applied to all the 
unpinned electrons thus preserving the equivalence 

\vspace{13.7 cm}

\noindent of the electrons with same 
number through the evolution. The parameters in ${\cal H}_{i(f)}$ used are:  
$N_i=N_f=61$, $M=75$, $z^{k,l}_i=0$, 
$z^{k,l}_f=z_0(\delta_{k,1}-\delta_{l,1})$.  The parameters $\vec r_i,\;
\;i=N_{i(f)}+1\;\;{\mathrm to}\;\; M$ can be inferred from Fig. 4.

PL spectra are computed for different recombination positions (the 
parameter $r_0$ 
in Eq. 1) of the electron numbered 1 (see Fig. 4). Each of the spectra 
is computed with $\Gamma_0$ being equal to the 
smallest phonon frequency of ${\cal H}_i^h$: $=3.46\times10^{-2}
\sqrt{e^2n^{1.5}/\epsilon_0m^*} $.  Fig. 5 shows their 
variation for recombination positions distributed uniformly along the edge of 
the Wigner-Seitz cell. Although the overall normalization 
increases rapidly with decreasing distance of the recombination center from 
the initial position of electron 1, the position of the peak shifts only 
slightly to the left. Hence the average of these spectra continues to have the 
typical width of each curve: $\sim 1/3$ of 
that  observed in time resolved PL. \cite{IVK1}  

The average PL spectrum - using 127 distinct points uniformly distributed 
in the Wigner-Seitz cell - is shown as an inset to Fig. 5. The spectrum 
is relatively narrow, and is in fact dominated by PL from recombination 
events in which the {\it x-y} motion of the electron is quite small. This is 
expected as it reflects the fact that the overlap between the recombining 
electron's initial and final states falls off rapidly as a function of its 
displacement. The spectrum is much narrower than seen in experiment, 
and does not reflect the experimentally observed double peak 
structure.\cite{IVK1,HB} The latter 
has been interpreted as evidence of liquid-solid coexistence due to disorder 
and/or finite temperatures,\cite{IVK1,HB,HA}, and our present results 
implicitly agree with this interpretation - a model of a perfect uniform 
solid as used in our work cannot explain the structure seen in continuous wave 
experiments.  

\begin{figure}
 \vbox to 8.5cm {\vss\hbox to 7cm
 {\hss\
   {\includegraphics{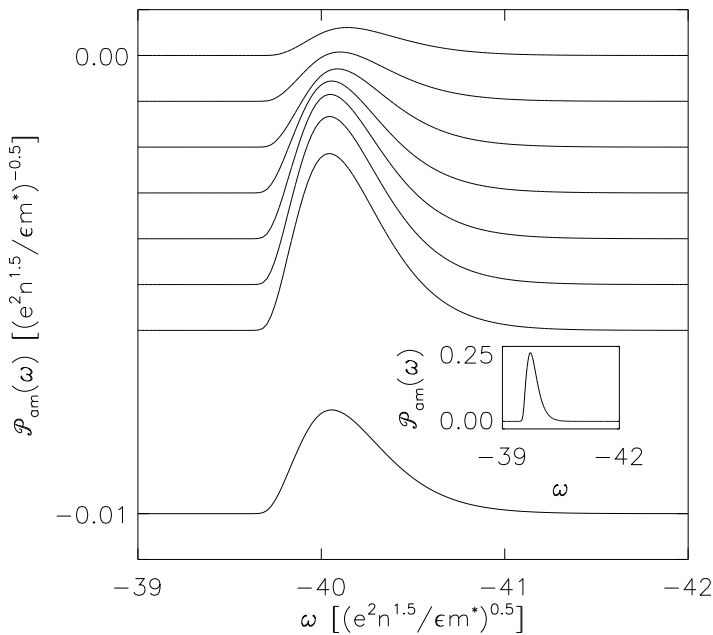}
   }
  \hss}
 }
\caption{ Spectra for recombination positions uniformly distributed along the 
edge of the 
Wigner-Seitz cell (see Fig. 4). They are shown in the order of decreasing 
distance from the initial position of the recombining electron. The final 
spectrum is the average of these spectra. It continues to have a width 
that is $\frac{\displaystyle 1}{\displaystyle 3}$ of that in experiment.$^8$ 
The inset shows the average spectrum corresponding to points distributed 
uniformly over the entire Wigner-Seitz cell (its axes are in the same units as 
the figure). It is dominated by the spectra 
corresponding to recombination positions close to the initial position of 
the recombining electron. }
\end{figure}

\subsection{Wigner lattice with disorder averaging}

We finally consider a disordered system (again with ideal Coulomb interactions) 
in which the disorder is due to 
pinned electrons below the {\it x-y} plane corresponding to 
charged acceptor atoms {\it i.e.} we assume that the spectrum of interest  
corresponds to a time when most of the recombination events have occurred so 
that all the acceptor atoms are charged in equilibrium. 
We consider 
a system size of 64 (initially) in-plane electrons. Unlike the previous cases, 
only these electrons are included in the computation of the density $n$. 
As the experimental density of acceptor atoms \cite{IVK1} is 
$5\times 10^9/$cm$^2$ the number of charged acceptor atoms corresponding to 
64 in-plane electrons must be very nearly 6. We therefore have 5 pinned 
electrons 
below the {\it x-y} plane corresponding to these atoms - the sixth would 
correspond to the recombination event for which the spectrum is computed.

In the absence any concrete information about the correlations between 
different acceptor 
atom coordinates that exist due to the process of $\delta$ doping 
we have assumed that each 
acceptor atom's {\it x-y} position is uniformly randomly distributed and is 
independent 
of any other acceptor atom's position. Therefore the 5 pinned electrons are 
independently placed randomly in the system. In principle, one would like to 
find the point farthest from all the electrons for a given disorder 
realization, and, since acceptors located near such points are least likely to
have a recombination event, assume that the late-time PL spectrum is 
dominated by an acceptor at this point. In practice, rather than generate a 
large number of disorder realizations, we choose just one, construct the 
Wigner-Seitz cells for the disordered system, and choose the corners of these 
cells as candidates for late-time PL recombination events. We believe this 
procedure will produce spectra qualitatively and even quantitatively close to 
that found by direct disorder-averaging, and is numerically much less time 
consuming. Essentially, we are assuming the system self-averages. To model 
the likelihood that a particular corner-configuration is likely to be 
available in the late-time spectroscopy, we additionally introduce a weighting 
factor for each corner. We do this by constructing the Voronoi cell 
around the lattice of corners, and set the weight for the corner to be 
proportional to its dual cell's area. Thus, acceptors that are particularly 
far from electrons are more likely to be available for PL events after 
long-time, and are given somewhat larger weights. 
   
Two cases are studied here: Case (a) - None of the 64 in-plane 
electrons are 
pinned, Case (b) - Two of the in-plane electrons that are maximally distant (as 
measured in the {\it x-y} plane) from the pinned electrons are additionally 
pinned {\it after} obtaining the initial equilibrium configuration (electrons 
numbered 13 and 47 in Fig. 6). The latter case is chosen to study what effects 
in-plane pinned electrons \cite{RMS} could have on the PL. As will be seen 
below, their effect is quite small. The parameters of 
${\cal H}_{i(f)}$ (Eq. 1) can therefore be set to:
$N_i=64$, $N_f=64$ for case (a) 

\clearpage

\noindent and $62$ for case (b), $M=69$, 
$z_i^{k,l}=z_0(\delta_{k,\geq 65}-\delta_{l,\geq 65})$, 
$z_f^{k,l}=z_0(\delta_{k,\geq 65}+\delta_{k,c}
-\delta_{l,\geq 65}-\delta_{l,c})$ 
where $c$ corresponds to the recombining electron. The parameters 
$\vec r_i,\;\;i=N_{i(f)}+1\;\;{\mathrm to}\;\; M$ can be inferred from Fig. 6.

The initial equilibrium configuration is reached through evolution 
beginning with a random distribution of the in-plane electrons. Beginning with 
this configuration, the final equilibrium configuration is reached in ten 
steps - each step localizing the recombining electron (through the 
choice of the parameter $\vec r_0$ in Eq. 1) closer to its final 
position which is chosen to be one of the corners of the Voronoi cell that 
initially contained the electron (see Fig. 6). This procedure 
allows us to find the closest minimum energy state in configuration space.

\widetext

\begin{figure}
 \vbox to 10.5cm {\vss\hbox to 18cm
 {\hss\
   {\includegraphics{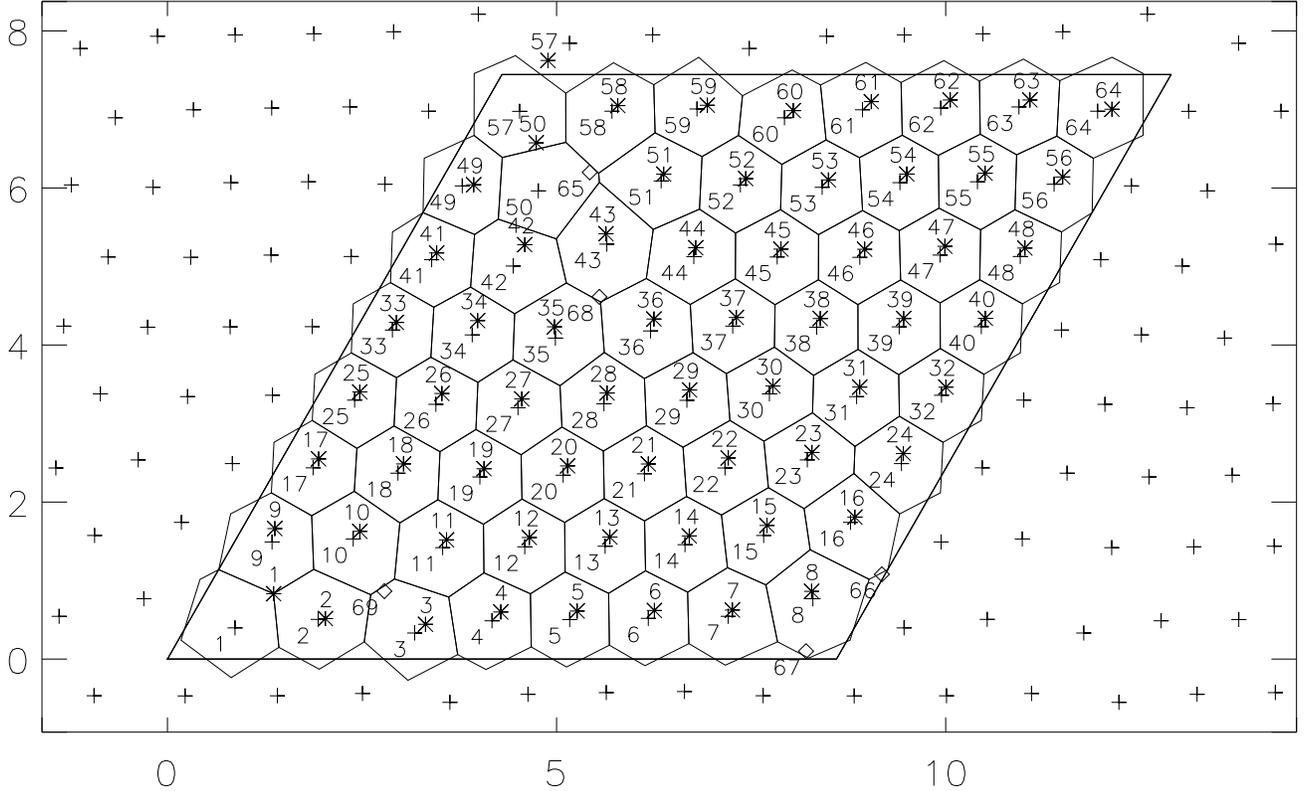}
   }
  \hss}
 }
\caption{ The figure shows the initial equilibrium configuration of unpinned 
electrons
(marked by $+$) and a particular final configuration of electrons (marked by 
$*$) corresponding to the recombination of the electron numbered 1 with a 
bound hole at a corner of the Voronoi cell that contained this electron. For 
easy visualization of the Voronoi cells, the initial configuration has been 
shown to be periodically repeated. Electrons that have already recombined are 
represented as pinned electrons (marked by $\diamond$ 
and numbered 65 through 69) that are independently and randomly placed in the 
system. They are not considered in the Voronoi construction. }
\end{figure}

\narrowtext

It appears that the width of the average spectra corresponding to the two 
cases studied is mainly due to the wide distribution ($D(\omega_p)$)  
of the position of the peak ($\omega_p$) in the individual spectra that 
constitute the 
average. This distribution is constructed by binning the peak 
positions into 
intervals in frequency. All peaks within a bin {\it do not} contribute 
equally to the distribution height - each contributes in proportion to 
the product of the 
corresponding Voronoi corner area and the area under 
the corresponding spectrum ($=\tilde {\cal P}_{am}(t=0)$). Fig. 7 
shows the 
normalized distributions for the two cases studied here. It 
may be seen that 
they have a width of $\sim 7 \sqrt{e^2n^{1.5}/\epsilon_0m^*}$. The 
insets show the corresponding distributions for a {\it second disorder 
realization} - they are similar in shape but have a somewhat smaller width: 
$\sim 5 \sqrt{e^2n^{1.5}/\epsilon_0m^*}$. We therefore expect that 
the average spectra corresponding to the first disorder realization to be 
roughly similar to the true disorder averaged spectra. 

The individual spectra in the average spectrum corresponding to the  
two cases studied here have been computed using $\Gamma_0 = 
3.80\times 10^{-2} \sqrt{e^2n^{1.5}/\epsilon_0 m^*}$ - the 
smallest non-zero phonon frequency of the perfect unpinned lattice with 
64 electrons. Fig. 8 shows these average

\vspace{13.7 cm}

\noindent spectra. As expected they are have a very large 
($> 7 \sqrt{e^2n^{1.5}/\epsilon_0m^*}$) width: $\sim 9 
\sqrt{e^2n^{1.5}/\epsilon_0m^*}$. This is $\sim 3$ times the 
experimental value.\cite{IVK1} The extra broadening seen in Fig. 8 over that of 
Fig. 7 is due to the shakeup of phonons. 

We speculate that the disagreement in these results and those found in 
experiment may be due to an overestimate of the disorder strength assumed in 
our uncorrelated random disorder model. 
In particular, the assumption that nearly all the holes on acceptor atoms have
recombined with electrons would be true only after very long waiting times, 
and it is unclear that this limit is actually achieved in experiment.\cite{IVK1}
Subsequent measurements \cite{IVK3} have suggested that it may indeed be the 
case that longer waiting times are needed. Clearly, one could ``tune'' 
the level of disorder in our model to obtain the experimentally observed 
linewidth.
However, this would require a painstaking and somewhat artificial 
``fine-tuning'' of our model. An independent 
measurement of the density of charged acceptors in these systems 
in the late time PL would greatly 
facilitate a quantitative comparison of this model with experiment. In this 
context it is worthwhile to point out that in general modulation $\delta$-doped 
heterostructures are known \cite{KD} to have substantial correlation among 
the impurity sites, and an uncorrelated random disorder model overestimates the 
disorder strength producing lower electron mobilities than experimentally 
observed. Any such correlation among the acceptor sites in our problem 
would reduce the spectral width of our calculated PL spectra, bringing 
experiment and theory into closer quantitative agreement.

\begin{figure}
 \vbox to 8.5cm {\vss\hbox to 7cm
 {\hss\
   {\includegraphics{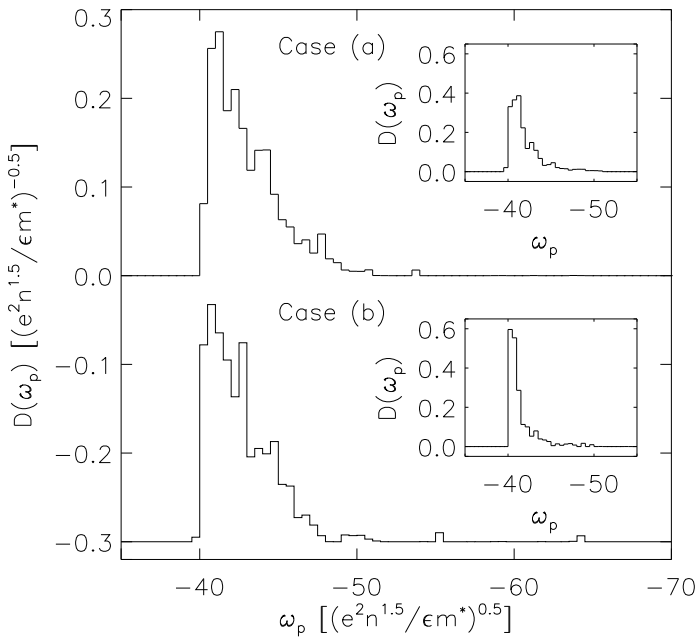}
   }
  \hss}
}
\caption{ The figure shows the distribution of peak positions (extended flat 
regions correspond to its zero value) in the spectra 
corresponding to the disorder realization shown in Fig. 6 for the two cases:
(a) with no in-plane pinning centers and (b) with two in-plane pinning centers. 
The broad distributions imply that the corresponding average spectra must be 
at least as wide.  The insets 
(whose axes are labeled in the same units as the figure) show the 
corresponding distributions for a second disorder realization. They 
are similar those of the first but have a slightly smaller width. This suggests 
that the two average spectra corresponding to the first disorder realization 
must be roughly similar to the true disorder averaged spectra.  }
\end{figure}
 
\begin{figure}
 \vbox to 8.5cm {\vss\hbox to 7cm
 {\hss\
   {\includegraphics{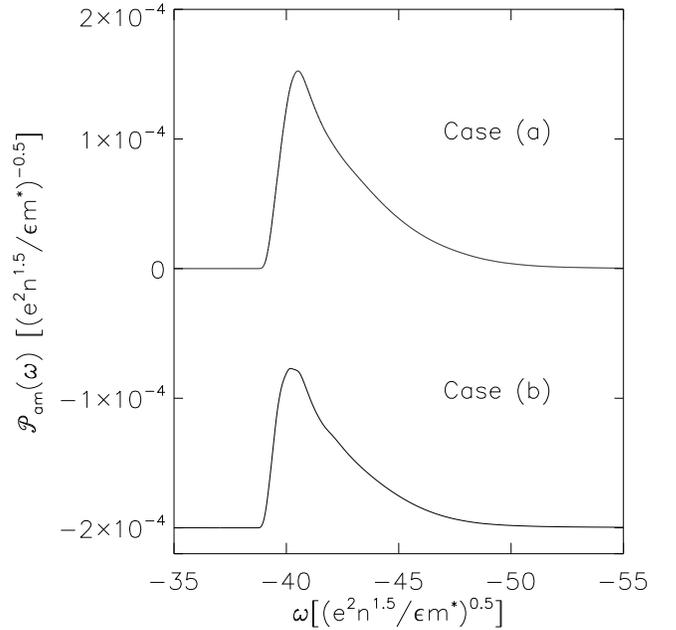}
   }
  \hss}
 }
\caption{  The figure shows the average spectra corresponding to the disorder 
realization shown in Fig. 6. Case (a) corresponds to all in-plane electrons 
being unpinned whereas case (b) corresponds to there being two additional 
in-plane pinning centers. As expected (see Fig. 7) the average spectra are 
very wide: roughly three times that observed by Kukushkin {\it et al}.$^8$}
\end{figure}

\section{Conclusion}

The line shape of the PL spectra due to electron recombination 
from a finite and pinned two dimensional Wigner crystal to a hole bound to an 
acceptor atom, computed using the harmonic approximation and first order time 
dependent perturbation theory, is similar to that seen in experiment
-it has a faster rising as compared to the falling edge. However, the 
width of the calculated spectra, for recombination events beginning with the 
perfect lattice configuration of electrons, is only the $\sim 1/3$ of that 
observed in experiment.
\cite{IVK1} With the initial configuration of electrons being disordered 
due to charged acceptor atoms corresponding to already recombined electrons, 
the spectral width is about three times the experimental value. We 
have speculated that considering lesser disordered configurations and 
recombination events different from those studied here may result in better 
agreement with experiment. Since disorder in our model is due only to 
charged acceptor atoms we speculate that any correlations in their
positions (that have been assumed to be zero here) may also reduce the 
corresponding computed spectral width. Because our perfect crystal and 
random disorder calculations 
give results factors of three smaller and larger than the experimental PL width
respectively, one could perhaps get {\it quantitative} agreement with 
the experimental spectra (we emphasize that our theoretical results are in 
good qualitative agreement with the 
experimental PL spectra) by using an adjustable correlated disorder model, but 
we feel that, in the absence of any concrete information about the nature of 
disorder, this is not a particularly meaningful exercise.

\section{Acknowledgments}   

This work has been supported by the U.S. ONR, the U.S. ARO, the NSF, and 
the Research Corporation. One of the authors 
(S. Kodiyalam) thanks Rodney Price, O.W. Greenberg, Lian Zheng, Alpan Raval 
and Mitrajit Dutta for useful discussions.

\appendix

\section{Ewalds sums}
The lattice sums appearing in the potential energy terms of Eq. 1 can be 
obtained as a special case of the following sum for $E$:

\begin{eqnarray}
 E(\vec r,\vec q,z,p,d)=
\displaystyle{\sum_{\displaystyle \vec R}} 
\frac{\displaystyle e^{\displaystyle i\vec q\cdot\vec R}}
{\displaystyle |\vec r + z\hat{e}_z + \vec R|^{\displaystyle p}} 
-\frac{ \displaystyle \theta(\vec r+z\hat e_z= 0)}
{\displaystyle |\vec r+z\hat{e}_z|^{\displaystyle p}}  \nonumber \\
-\frac{ \displaystyle \theta(p\leq d,\vec q=0)}{\displaystyle A} 
{\displaystyle \int}
\frac{\displaystyle d^dr \, ^\prime e^{\displaystyle i\vec q\cdot
\vec r \, ^\prime}}{\displaystyle |\vec r + z\hat{e}_z +
\vec r \, ^\prime |^{\displaystyle p}}  \;, \\
 {\mathrm with} \;\; \vec R=\sum_{\displaystyle i=1}^
{\displaystyle d}  m_i\vec a_i  , \; \;
m_i  \in {\mathrm integers}, \; \; 
\vec a_i\cdot\hat{e}_z=0 \forall \; i \;.  \nonumber 
\end{eqnarray}
\noindent In the above formula $\vec a_i$ are the primitive lattice 
vectors of the $d$ dimensional direct space lattice, and $A$ is the unit cell 
volume.
$\vec r$ is assumed to lie $within$ the region spanned by the
unit cells touching the origin of the direct lattice.
Similarly, $\vec q$ is assumed to lie $within$ the region spanned by the
unit cells touching the origin of the reciprocal lattice, the unit cell of 
which is defined by the vectors  $\vec b_i$ that are determined by the 
following:
\begin{eqnarray}
\vec a_i\cdot\vec b_j =2\pi\delta_{ij} \;. \nonumber 
\end{eqnarray}
\noindent The unit vector $\hat{e}_z$ in Eq. A1 does not belong to the space 
spanned by $\vec a_i$  and lies along an additional dimension ``perpendicular''
 to the direct space 
lattice. $p$ may be any positive real number. The function $\theta$  is equal 
to one if all the conditions in its argument are true and zero otherwise. 
In the Madelung sum, we are interested in the interaction energy of a 
particle with all other particles, including its images in other unit cells. 
This is most conveniently handled by a sum over the Bravais lattice vectors 
(first term in Eq. A1), but then one must remove the interaction of a particle 
with itself (second term in Eq. A1). For long rance interactions ($p\geq d$), 
such sums diverge unless an interaction with a neutralizing background is 
introduced. This is the meaning of the last term in Eq. A1. $\tilde{V}$ that 
appears in Eq. 1 may now be written as: 
\begin{eqnarray}
\tilde{V}(\vec r,z) = \frac{\displaystyle e^2}{\displaystyle \epsilon_0}
E(\vec r,0,z,1,2) \;. \nonumber
\end{eqnarray}

Each term of the lattice sum ($LS$) (the summation in Eq. A1) for $E$ can 
be written as a sum of a rapidly 
decreasing function $f_1(\vec R)$  and another function $f_2(\vec R)$ that is 
slowly varying. This division is achieved by multiplying the terms by the 
following expression for unity in terms of gamma functions:

\begin{eqnarray}
 1=\left[ \Gamma_{\frac{\displaystyle p}{\displaystyle 2}} \right]^{-1}
\left[ \Gamma( \frac{p}{2},\epsilon^2 x^2) + 
       \gamma( \frac{p}{2},\epsilon^2 x^2) \right] \;, \nonumber 
\end{eqnarray}
\noindent where
\begin{eqnarray}
x=|\vec r+z\hat{e}_z+\vec R|\;\;,\;\;\Gamma(a,b)={\displaystyle \int}_ 
{\displaystyle b}^ {\displaystyle \infty} e^{ -t} t^{a-1}dt \;,\nonumber\\ 
\Gamma_{a}=\Gamma(a,0)\;\;\;\;\;,\;\;\;\; \gamma(a,b) = \Gamma_a - \Gamma(a,b)  
\; , \nonumber
\end{eqnarray}

\noindent and $\epsilon$ is a suitably chosen (inverse length scale) 
convergence
parameter the choice of which is explained later in this section.
Hence the functions $f_1$ and $f_2$  may be identified as:

\begin{eqnarray}
f_1(\vec R)=\Gamma_{\frac{\displaystyle p}{\displaystyle 2}}^{-1}
x^{\displaystyle -p} 
e^{\displaystyle i\vec q \cdot \vec R}
\; \Gamma( \frac{p}{2},\epsilon^2x^2) \;, \\
f_2(\vec R)=\Gamma_{\frac{\displaystyle p}{\displaystyle 2}}^{-1} 
x^{\displaystyle -p} 
e^{\displaystyle i\vec q \cdot \vec R}
\; \gamma( \frac{p}{2},\epsilon^2x^2) \;, \nonumber
\end{eqnarray}

\noindent with the $LS$ given by

\begin{eqnarray}
LS = \sum_{\displaystyle \vec R} f_1(\vec R) + 
     \sum_{\displaystyle \vec R} f_2(\vec R)  \;. 
\end{eqnarray}

Due to the functional form of the gamma function $\Gamma(a,b)$,
the terms of the first summation in Eq. A3 die off rapidly with 
increasing $\epsilon|\vec R|$ {\i.e.} for $\epsilon|\vec R| \gg 1$,

\begin{eqnarray}
|f_1(\vec R)| \sim 
\Gamma_{\frac{\displaystyle p}{\displaystyle 2}}^{-1}\epsilon^{\displaystyle p} 
(\epsilon|\vec R|)^{-2}e^{\displaystyle -(\epsilon|\vec R|)^2} \;. 
\end{eqnarray} 

\noindent Therefore this summation is carried out 
directly over \{$\vec R$\} (beginning with $\vec R=0$) until the estimated 
$relative$ error due to the neglected terms is smaller than a chosen value.  

Due to the functional form of the gamma function $\gamma(a,b)$, the 
terms of the second summation in Eq. A3 vary slowly with increasing 
$\epsilon|\vec R|$.  Therefore their summation is better performed using the 
following identity that transforms the summation over \{$\vec R$\} of $f_2$ to 
a summation over reciprocal lattice vectors (\{$\vec G$\}) of the Fourier 
transform of $f_2$ ($\tilde f_2$):
\begin{eqnarray}
\sum_{\displaystyle \vec R} f_2(\vec R) = \int f_2(\vec r \,^\prime)
\delta_p(\vec r\,^\prime) d^d r\,^\prime \;, \; \; \delta_p(\vec r\,^\prime)  = 
\sum_{\displaystyle \vec R} &\delta(\vec r\,^\prime - \vec R) \;. \nonumber \\
\Rightarrow \sum_{\displaystyle \vec R} f_2(\vec R) = \frac{\displaystyle 1}
{\displaystyle (2\pi)^d} \int \tilde f_2(\vec k) 
\tilde \delta_p(-\vec k) d^d k \;, & \nonumber
\end{eqnarray}

\begin{eqnarray}
{\mathrm where} \; \; \; \tilde f_2(\vec k)&=&\int f_2(\vec r\,^\prime) 
e^{\displaystyle i\vec k\cdot \vec r\,^\prime} d^d r\,^\prime \;,  \nonumber \\
 \tilde \delta_p(\vec k) &=& \int \delta_p(\vec r\,^\prime) 
e^{\displaystyle i\vec k\cdot \vec r\,^\prime} d^d r\,^\prime 
 = \frac{\displaystyle (2\pi)^d}{\displaystyle A} \displaystyle{ \sum_{\vec G}} 
\delta(k-\vec G) \;, \nonumber \\ 
{\mathrm with} \;\;\;\vec G&=& \sum_{\displaystyle i=1}^{\displaystyle d} 
m_i\vec b_i \;, \; \;  m_i \in {\mathrm integers,\;so\;that}  \nonumber \\
\sum_{\displaystyle \vec R} f_2(\vec R)&=& \frac{\displaystyle 1}
{\displaystyle A} {\displaystyle \sum_{\vec G}} \tilde f_2(-\vec G)\;.\nonumber 
\end{eqnarray}
\noindent $\tilde f_2(\vec k)$ is computed to be: 
\begin{eqnarray}
\tilde f_2(\vec k) = 
\frac{ \Gamma_{\frac{\displaystyle p}{\displaystyle 2}}^{-1}
\displaystyle \epsilon^{\displaystyle p}}
{\displaystyle (\epsilon/\sqrt{\pi})^{\displaystyle d}}
e^{\displaystyle -i(\vec k + \vec q)\cdot \vec r}
\Upsilon(\frac{\displaystyle d-p}{\displaystyle 2},
\left | \frac{\displaystyle \vec k + \vec q}{\displaystyle 2\epsilon} \right|^2
,\epsilon^2 z^2) \;, \nonumber \\ 
 {\mathrm where} \;\; \; \Upsilon(a,b,c) = {\displaystyle \int}
_{\displaystyle 1}^{\displaystyle \infty}e^{-(\frac{c}{t} + bt)} t^{a-1}dt \;,
{\mathrm and \;\; hence}  \nonumber 
\end{eqnarray}
\begin{eqnarray}
\sum_{\displaystyle \vec R} f_2(\vec R)&=&{\displaystyle \sum_{\vec G} } 
\frac{ \Gamma_{\frac{\displaystyle p}{\displaystyle 2}}^{-1}
\displaystyle \epsilon^{\displaystyle p}}
{\displaystyle A(\epsilon/\sqrt{\pi})^{\displaystyle d}}
\times   \nonumber \\
&& e^{\displaystyle i(\vec G - \vec q)
\cdot \vec r} 
\Upsilon(\frac{\displaystyle d-p}{\displaystyle 2},
\left ( \frac{\displaystyle y}{\displaystyle 2\epsilon} \right)^2
,\epsilon^2 z^2) \;, \\ 
 {\mathrm with} \;\;\; y &=& |\vec G - \vec q| \;.   \nonumber
\end{eqnarray}

It must be noted that $\Upsilon(a,b,c) \leq \Upsilon(a,b,0)=
b^{-a}\Gamma(a,b)$. Therefore, again due to the functional 
form of the gamma function $\Gamma(a,b)$, the terms in the second summation of 
Eq. A5 die off rapidly with increasing $\frac{\displaystyle |\vec G|}
{\displaystyle 2\epsilon}$ {\it i.e.} for $\frac{\displaystyle |\vec G|}
{\displaystyle 2\epsilon} \gg 1$
\begin{eqnarray}
\frac{\displaystyle |\tilde f_2(-\vec G)|}{A} \stackrel{\displaystyle <}
{\displaystyle \sim} 
\frac{ \Gamma_{\frac{\displaystyle p}{\displaystyle 2}}^{-1} 
\displaystyle \epsilon^{\displaystyle p}}
{\displaystyle A(\epsilon/\sqrt{\pi})^{\displaystyle d}}
\left( \frac{\displaystyle |\vec G|}{\displaystyle 2\epsilon} \right)^{-2} 
e^{\displaystyle -\left( \frac{|\vec G|}{2\epsilon} \right)^2} \;.
\end{eqnarray}  

\noindent Therefore the second summation in Eq. A3 is transformed 
using A5 and is carried out over $\{\vec G\}$ (beginning with $\vec G=0$) 
until 
the estimated $relative$ error due to the neglected terms is smaller than a 
chosen value. 

$E$  can now be written (using Eqs. A1,A2,A3 and A5) as:
\begin{eqnarray}
\lefteqn{E(\vec r,\vec q,z,p,d)= }\nonumber \\
&\Gamma_{\frac{\displaystyle p}{\displaystyle 2}}^{-1}\epsilon^{\displaystyle p}
{\displaystyle \sum_{\displaystyle \vec R}} 
\frac{ \displaystyle e^{\displaystyle i\vec q\cdot\vec R}
\Gamma(\frac{\displaystyle p}{\displaystyle 2},\epsilon^2
|\vec r +z\hat e_z+\vec R|^2)}
{\displaystyle (\epsilon|\vec r +z\hat e_z+\vec R|)^{\displaystyle p}}
-\frac{ \displaystyle \theta(\vec r+z\hat e_z= 0)}
{\displaystyle |\vec r+z\hat{e}_z|^{\displaystyle p}} \nonumber \\
&+\frac{ \displaystyle \Gamma_{\frac{\displaystyle p}{\displaystyle 2}}^{-1}
\displaystyle \epsilon^{\displaystyle p}}{\displaystyle A(\epsilon/\sqrt{\pi})^
{\displaystyle d}  }
{\displaystyle \sum_{\displaystyle \vec G}} 
e^{\displaystyle i(\vec G - \vec q) \cdot \vec r}
\Upsilon(\frac{\displaystyle d-p}{\displaystyle 2},
\left ( \frac{\displaystyle |\vec G - \vec q|}{\displaystyle 2\epsilon} 
\right)^2 ,\epsilon^2 z^2) \nonumber \\ 
&-\frac{ \displaystyle \theta(p\leq d,\vec q=0)}{\displaystyle A}
{\displaystyle \int}
\frac{\displaystyle d^dr \, ^\prime e^{\displaystyle i\vec q\cdot
\vec r \, ^\prime}}{\displaystyle |\vec r + z\hat{e}_z +
\vec r \, ^\prime |^{\displaystyle p}} \;. \nonumber
\end{eqnarray}
\noindent It may be rewritten, separating the possibly singular terms 
(corresponding to $\vec R=0$ and $\vec G=0$ ) and evaluating 
them as appropriate limiting cases (of $\vec r+z\hat e_z\rightarrow 0$ and 
$\vec q \rightarrow 0$ respectively) as:
\begin{eqnarray}
\lefteqn{E(\vec r,\vec q,z,p,d)= \;\;\;\;\;\;\;\;\;\;\;\;\;\;\;\;\;\;\;\;   
\;\;\;\;\;\;\;\;\;\;\;\;\;\;\;\;\;\;\;\;\;\;\;\;\;\;\;\;\;\;\;\;\;\;\;
{\mathrm (A7)} }\nonumber\\
&\Gamma_{\frac{\displaystyle p}{\displaystyle 2}}^{-1}\epsilon^{\displaystyle p}
\left[x+{\displaystyle \sum_{\displaystyle \vec R \neq 0}}
\frac{ \displaystyle e^{\displaystyle i\vec q\cdot\vec R}
\Gamma(\frac{\displaystyle p}{\displaystyle 2},\epsilon^2
|\vec r +z\hat e_z+\vec R|^2)}
{\displaystyle (\epsilon|\vec r +z\hat e_z+\vec R|)^{\displaystyle p}} \right] 
\nonumber \\
\lefteqn { \left. 
+\frac{ \displaystyle \Gamma_{\frac{\displaystyle p}{\displaystyle 2}}^{-1}
\displaystyle \epsilon^{\displaystyle p}}{\displaystyle A(\epsilon/\sqrt{\pi})^
{\displaystyle d}  }
\right[ y e^{\displaystyle -i\vec q\cdot \vec r}  \; + } \nonumber \\
& \left. \;\;\;\;\;\;\;\;{\displaystyle \sum_{\displaystyle \vec G \neq 0}}
e^{\displaystyle i(\vec G - \vec q) \cdot \vec r}
\Upsilon(\frac{\displaystyle d-p}{\displaystyle 2},
\left ( \frac{\displaystyle |\vec G - \vec q|}{\displaystyle 2\epsilon}
\right)^2 ,\epsilon^2 z^2) \right] \;, \nonumber 
\end{eqnarray}
\noindent with
\begin{eqnarray}
x&=&\left\{
\begin{array}{ll}
\frac{\displaystyle \Gamma(\frac{\displaystyle p}{\displaystyle 2},\epsilon^2,
|\vec r +z\hat e_z|^2)}{\displaystyle \epsilon|\vec r+z\hat e_z|^
{\displaystyle p}} \;\;\;\;\;\;\;\;\;\; & 
\mbox{if $\vec r + z\hat e_z \neq 0$} \\
-\frac{\displaystyle 2}{\displaystyle p} &\mbox{if $\vec r + z\hat e_z = 0$} 
\end{array}
\right. \nonumber \\
y&=&\left\{
\begin{array}{ll}
\Upsilon(\frac{\displaystyle d-p}{\displaystyle 2},
\left( \frac{\displaystyle |-\vec q |}{\displaystyle 2\epsilon} \right)^2,
\epsilon^2 z^2)
& \mbox{if $p > d$ or $\vec q \neq 0$} \\
-\Upsilon(\frac{\displaystyle p-d}{\displaystyle 2},\epsilon^2 z^2,0)
&\mbox{if $p \leq d$ and $\vec q = 0 \;. $}
\end{array}
\right. \nonumber
\end{eqnarray}

\addtocounter{equation}{1}

Equation A7 can be  used to evaluate $E$ for all values of its arguments. 
First and second derivatives of $E$ (with respect to $\vec r$) that are needed 
to compute the forces and the dynamical matrix are also obtained by 
performing the requisite operations on the above expression. 
For {\it equally rapid} convergence of the two summations in this equation, 
$\epsilon$ must be given by:
\begin{eqnarray}
\epsilon = \frac{\displaystyle \sqrt{\pi}}{\displaystyle A^{\displaystyle 1/d}}
\; .
\end{eqnarray}
\noindent With this choice of $\epsilon$, the magnitudes of the largest
term neglected ($|\delta T|$), if $N$ of the largest terms are included in 
each of the two summations, become equal - and may be estimated 
(using Eqs. A4 and A6) to be:  
\begin{eqnarray}
|\delta T|=\Gamma_{\frac{\displaystyle p}{\displaystyle 2}}^{-1} 
\left(    \frac{\displaystyle \sqrt{\pi}}{\displaystyle A^
{\displaystyle 1/d} } \right)^{\displaystyle p} \times 
\xi^{-2} e^{\displaystyle -\xi^2} \;,\;\;\;\xi=\left(
\Gamma_{\frac{\displaystyle d}{\displaystyle 2}+1} N \right)^
{\displaystyle 1/d} \; .\nonumber
\end{eqnarray}

From previous work,\cite{VM} $\Upsilon(a,b,c)$ can be written in terms of 
Gamma functions for $a$ being any odd multiple of $\frac{1}{2}$ (and $b\neq 0$) 
using: 
\begin{eqnarray}
\Upsilon(a,b,c) = \frac{\displaystyle 1}{\displaystyle b}\left[ 
e^{\displaystyle -(b+c)}+ a\Upsilon(a-1,b,c)+c\Upsilon(a-2,b,c) \right]
\;, \nonumber \\
\Upsilon(\frac{1}{2},b,c)=\frac{\displaystyle 1}{\displaystyle 2\sqrt{b}} 
\left[\alpha + \beta \right]\;\;\;,\;\;\;
\Upsilon(-\frac{1}{2},b,c)=\frac{\displaystyle 1}{\displaystyle 2\sqrt{c}}
\left[\alpha - \beta \right] \; , \nonumber 
\end{eqnarray}
\noindent with 
\begin{eqnarray}
\alpha &=& e^{\displaystyle -2\sqrt{bc}}\left( \Gamma_{\frac{
\displaystyle 1}{\displaystyle 2}}  - 
\left[ \theta(b\geq c) - \theta(b \leq c) \right ] \gamma(\frac{\displaystyle 1}
{\displaystyle 2},|\sqrt{b}-\sqrt{c}|) \right ) \; , \nonumber \\
\beta &=& e^{\displaystyle 2\sqrt{bc}} \Gamma(\frac{\displaystyle 1}
{\displaystyle 2},\sqrt{b}+\sqrt{c}) \; . \nonumber 
\end{eqnarray}
\noindent Finally, an accurate evaluation of Gamma functions 
({\it relative error} 
$\stackrel{\displaystyle <}{\sim} 10^{-14}$) is done using the ideas 
outlined by C. Lanczos. \cite{Lz,NR}

\section{Normal modes of a general harmonic Hamiltonian} 

This section shows the procedure for transforming a harmonically approximated 
Hamiltonian of the form   
\begin{eqnarray}
\lefteqn{  {\cal H}^h= \epsilon^m + }\nonumber \\
&\begin{array}{ccc}
  \frac{1}{2m^*}\left[{\mathbf \Phi}^T {\mathbf \Pi}^T \right] &
  \left[ \begin{array}{cc}
          m^*{\mathbf D}+{\mathbf B}^T{\mathbf B} & -{\mathbf B}^T \\
           -{\mathbf B}                            & {\mathbf I}
         \end{array}
  \right] &
  \left[ \begin{array}{c}
           {\mathbf \Phi} \\
           {\mathbf \Pi}
         \end{array}
  \right]
                \end{array}  \; \nonumber
\end{eqnarray}
\noindent into a summation over uncoupled (normal) modes by a linear 
canonical transformation of the phase space variables 
$\left[{\mathbf \Phi}^T {\mathbf \Pi}^T \right]$  (Eqs. 4 and 5).
The only requirement for this procedure to apply is that ${\mathbf D}$ 
be a symmetric positive-definite matrix.
This is the case here since
${\mathbf D}$ is the dynamical matrix corresponding to a {\it stable} 
equilibrium. (All matrices that appear in this section are real.)
The transformation is carried out in three steps which together constitute  a 
canonical transformation. This is verified by simultaneously determining the 
equations of motion for the transformed variables and showing that they may be 
derived from the Hamiltonian in the same variables. The equations of 
motion in the current variables read as:
\begin{eqnarray}
\left[
\begin{array}{c}
{\mathbf \dot{\Phi}} \\
{\mathbf \dot{\Pi}}
\end{array} 
\right] = \frac{\displaystyle 1}{\displaystyle m^*}
\left[
\begin{array}{cc}
-{\mathbf B} & {\mathbf I} \\
-m^*{\mathbf D}-{\mathbf B}^T{\mathbf B} & {\mathbf B}^T 
\end{array}
\right]
\left[
\begin{array}{c}
{\mathbf \Phi} \\
{\mathbf \Pi}
\end{array}
\right]  \; . \nonumber
\end{eqnarray}
\noindent The procedure is similar to the standard technique,\cite{Sky} 
but has to be extended to 
a multiparticle generalization.

The first of the three stpes in the transformation
is in itself a canonical transformation that diagonalizes the dynamical 
matrix ${\mathbf D}$. This is accomplished by an orthogonal matrix 
${\mathbf N}$ ({\it i.e.} ${\mathbf N}^T={\mathbf N}^{-1}$) the existence of 
which follows from the real symmetric structure of ${\mathbf D}$. 
Since the eigenvalues of ${\mathbf D}$ are all positive (as it is a 
positive definite matrix) the diagonalized 
result may be written as: $m^*{\mathbf D}_\omega^2={\mathbf N}^T {\mathbf D} 
{\mathbf N}$ where ${\mathbf D}_\omega$ is a diagonal matrix with positive 
elements of the dimension of frequency. The transformed variables may 
therefore be written as:
\begin{eqnarray}
\lefteqn{ 
\left[
\begin{array}{c}
{\mathbf \Phi_1} \\
{\mathbf \Pi_1}
\end{array}
\right] = 
          } \\
&\left[
\begin{array}{cc}
\left(m^*{\mathbf D}_\omega\right)^{1/2}{\mathbf N}^T & {\mathbf 0} \\
  {\mathbf 0}    & \left(m^*{\mathbf D}_\omega\right)^{-1/2}{\mathbf N}^T
\end{array}
\right]
\left[
\begin{array}{c}
{\mathbf \Phi} \\
{\mathbf \Pi}
\end{array}
\right] \;, \nonumber
\end{eqnarray}
\noindent in terms of which the Hamiltonian and the equations of motion 
read:
\begin{eqnarray}
\lefteqn{  {\cal H}^h= \epsilon^m + }\nonumber \\
&\begin{array}{ccc}
  \frac{1}{2}\left[{\mathbf \Phi_1}^T {\mathbf \Pi_1}^T \right] &
  \left[ \begin{array}{cc}
          {\mathbf D}_\omega+{\mathbf B}_\omega^T{\mathbf D}_\omega^{-1}
                             {\mathbf B}_\omega & -{\mathbf B}_\omega^T \\
           -{\mathbf B}_\omega                  & {\mathbf D}_\omega
         \end{array}
  \right] &
  \left[ \begin{array}{c}
           {\mathbf \Phi_1} \\
           {\mathbf \Pi_1}
         \end{array}
  \right]
                \end{array}  \;, \nonumber
\end{eqnarray}
\begin{eqnarray}
\left[
\begin{array}{c}
{\mathbf \dot{\Phi}_1} \\
{\mathbf \dot{\Pi}_1}
\end{array}
\right]& = 
\left[
\begin{array}{cc}
-{\mathbf B}_\omega & {\mathbf D}_\omega \\
-{\mathbf D}_\omega-{\mathbf B}_\omega^T{\mathbf D}_\omega^{-1}
 {\mathbf B}_\omega & {\mathbf B}_\omega^T
\end{array}
\right]
\left[
\begin{array}{c}
{\mathbf \Phi_1} \\
{\mathbf \Pi_1}
\end{array}
\right] \;, \nonumber
\end{eqnarray}
\noindent where ${\mathbf B}_\omega$ is the transformed ${\mathbf B}$ matrix 
that also has the dimensions of frequency and is given 
by: $m^*{\mathbf B}_\omega={\mathbf D}_\omega^{-1/2}{\mathbf N}^T 
{\mathbf B}{\mathbf N}{\mathbf D}_\omega^{1/2}$. It may be seen that if 
${\mathbf B}={\mathbf 0}$ ($\Rightarrow {\mathbf B}_\omega={\mathbf 0}$) 
then Eq. B1 would suffice to diagonalize ${\cal H}^h$ into the required 
form. 

The next transformation trivially diagonalizes the Hamiltonian. However, as the 
equations of motion in the new variables do not follow from the Hamiltonian in 
the same variables, it is not a canonical transformation. The transformation 
is given by:
\begin{eqnarray}
\left[
\begin{array}{c}
{\mathbf \Phi_2} \\
{\mathbf \Pi_2}
\end{array}
\right] =
\left[
\begin{array}{cc}
{\mathbf D}_\omega^{1/2} & {\mathbf 0} \\
-{\mathbf D}_\omega^{-1/2}{\mathbf B}_\omega  & {\mathbf D}_\omega^{1/2}
\end{array}
\right]
\left[
\begin{array}{c}
{\mathbf \Phi_1} \\
{\mathbf \Pi_1}
\end{array}
\right] \;.
\end{eqnarray}
\noindent In terms of the new variables the Hamiltonian and the equations 
of motion now read as: 
\begin{eqnarray}
 {\cal H}^h&=& \epsilon^m + 
\begin{array}{ccc}
  \frac{1}{2}\left[{\mathbf \Phi_2}^T {\mathbf \Pi_2}^T \right] &
  \left[ \begin{array}{cc}
            {\mathbf I} & {\mathbf 0} \\
            {\mathbf 0} & {\mathbf I}
         \end{array}
  \right] &
  \left[ \begin{array}{c}
           {\mathbf \Phi_2} \\
           {\mathbf \Pi_2}
         \end{array}
  \right]
                \end{array}  \;, \nonumber \\
\left[
\begin{array}{c}
{\mathbf \dot{\Phi}_2} \\
{\mathbf \dot{\Pi}_2}
\end{array}
\right]& =&
\left[
\begin{array}{cc}
{\mathbf 0} & {\mathbf D}_\omega \\
-{\mathbf D}_\omega & {\mathbf \tilde{B}}_\omega^T - {\mathbf \tilde{B}}_\omega
\end{array}
\right]
\left[
\begin{array}{c}
{\mathbf \Phi_2} \\
{\mathbf \Pi_2}
\end{array}
\right] \nonumber \\
& =& {\mathbf Q} 
\left[
\begin{array}{c}
{\mathbf \Phi_2} \\
{\mathbf \Pi_2}
\end{array}
\right] \;, \nonumber
\end{eqnarray}
\noindent where ${\mathbf \tilde{B}}_\omega={\mathbf D}_\omega^{-1/2}
{\mathbf B}_\omega{\mathbf D}_\omega^{1/2}$.

The final transformation, which together with the previous one results in 
a canonical transformation, involves the ``block off-diagonalization'' of 
the matrix ${\mathbf Q}$ by an orthogonal matrix ${\mathbf M}$ ({\it i.e.} 
${\mathbf M}^t={\mathbf M}^{-1}$) the existence of which follows from 
the real antisymmetric structure of ${\mathbf Q}$. It must be noted that 
${\mathbf Q}$ is an even-dimensional matrix ($2N_f\times 2N_f$) which may be 
shown to be non-singular since all the diagonal elements of 
${\mathbf D}_\omega$ are non-zero. ${\mathbf M}$ is constructed  
using the eigenvectors of another {\it symmetric} matrix 
$\acute{\mathbf {Q}}$:
\begin{eqnarray}
\acute{\mathbf {Q}}=
\left[
\begin{array}{cc}
    {\mathbf 0} & {\mathbf Q} \\
   -{\mathbf Q} & {\mathbf 0} 
\end{array} 
\right] \;. \nonumber
\end{eqnarray} 
\noindent It can be shown that the eigenvalues of $\acute{\mathbf {Q}}$ 
are necessarily non-zero (due to ${\mathbf Q}$ being nonsingular), degenerate 
by an even number and that corresponding to 
every eigenvalue there is another with equal magnitude and opposite sign. 
This can be seen from the construction of the eigenvectors: if there is 
an eigenvector of the form $[{\mathbf v_1}^T,{\mathbf v_2}^T]^T$ with 
eigenvalue $\omega_1$ (${\mathbf v_1}$ and ${\mathbf v_2}$ are column 
vectors with $2N_f$ components), then the vectors 
$[{\mathbf v_2}^T,-{\mathbf v_1}^T]^T$, $[{\mathbf v_2}^T,{\mathbf v_1}^T]^T$ 
and $[-{\mathbf v_1}^T,-{\mathbf v_2}^T]^T$ are also eigenvectors with 
eigenvalues $\omega_1$, $-\omega_1$ and $-\omega_1$. It may be shown that 
the four vectors are mutually orthogonal and hence linearly independent. 
From the mutual orthogonality of all eigenvectors of $\acute{\mathbf {Q}}$ 
constructed in the above fashion (for eigenvalues that are more than two 
fold degenerate, the corresponding  eigenvectors may need an explicit 
orthogonalization procedure) it can be shown that the vectors ${\mathbf v_1}^i$,
${\mathbf v_2}^i$ ($i=1$ to $N_f$) are mutually orthogonal. They 
are chosen from the set of eigenvectors of ${\mathbf \tilde{Q}}$ that 
have $\omega_i>0$ and are explicitly normalized. The matrix 
${\mathbf M}$ is then constructed: 
\begin{eqnarray}
{\mathbf M}=[{\mathbf v_1}^1,{\mathbf v_1}^2,.....,{\mathbf v_1}^{N_f},
{\mathbf v_2}^1,{\mathbf v_2}^2,.....,{\mathbf v_2}^{N_f}]  \;. \nonumber
\end{eqnarray}
\noindent It may be verified that ${\mathbf M}$ is orthogonal and that it 
``block off-diagonalizes'' ${\mathbf Q}$:
\begin{eqnarray}
{\mathbf M}^T{\mathbf Q}{\mathbf M} = 
\left[
\begin{array}{cc}
  {\mathbf 0} & {\mathbf \Omega} \\
 -{\mathbf \Omega} & {\mathbf 0}  
\end{array} \right] \;, \nonumber
\end{eqnarray}
\noindent where ${\mathbf \Omega}$ is a diagonal matrix with the diagonal 
elements being the (positive) phonon frequencies $\omega_i$. The final 
transformation is given by:
\begin{eqnarray}
\left[
\begin{array}{c}
{\mathbf \Psi} \\
{\mathbf \Xi}
\end{array}
\right] =
\left[
\begin{array}{cc}
{\mathbf \Omega}^{-1/2} & {\mathbf 0} \\
{\mathbf 0}  & {\mathbf \Omega}^{-1/2} 
\end{array}
\right]
\left[ {\mathbf M}^T \right]
\left[
\begin{array}{c}
{\mathbf \Phi_2} \\
{\mathbf \Pi_2}
\end{array}
\right]  \;. 
\end{eqnarray}
\noindent In terms of the variables ${\mathbf \Psi}$ and ${\mathbf \Xi}$ the 
the Hamiltonian and the equations of motion read: 
\begin{eqnarray}
 {\cal H}^h&=& \epsilon^m +
\begin{array}{ccc}
  \frac{1}{2}\left[{\mathbf \Psi}^T {\mathbf \Xi}^T \right] &
  \left[ \begin{array}{cc}
            {\mathbf \Omega} & {\mathbf 0} \\
            {\mathbf 0} & {\mathbf \Omega}
         \end{array}
  \right] &
  \left[ \begin{array}{c}
           {\mathbf \Psi} \\
           {\mathbf \Xi}
         \end{array}
  \right]
                \end{array}  \;, \nonumber \\
\left[
\begin{array}{c}
{\mathbf \dot{\Psi}} \\
{\mathbf \dot{\Xi}}
\end{array}
\right]& =&
\left[
\begin{array}{cc}
{\mathbf 0} & {\mathbf \Omega} \\
-{\mathbf \Omega} & {\mathbf 0}
\end{array}
\right]
\left[
\begin{array}{c}
{\mathbf \Psi} \\
{\mathbf \Xi}
\end{array}
\right] \;. \nonumber 
\end{eqnarray}
\noindent As the equations of motion for ${\mathbf \Psi}$ and ${\mathbf \Xi}$ 
follow from the Hamiltonian in the same variables, the combined transformation
${\mathbf C}$ represented by Eqs. B1, B2 and B3 must be canonical. In 
other words the transformation ${\mathbf C}$ given by: 
\begin{eqnarray}
{\mathbf C}&=&
\left[
\begin{array}{cc}
{\mathbf \Omega}^{-1/2} & {\mathbf 0} \\
{\mathbf 0}  & {\mathbf \Omega}^{-1/2}
\end{array}
\right]
\left[ {\mathbf M}^T \right]
\left[
\begin{array}{cc}
{\mathbf D}_\omega^{1/2} & {\mathbf 0} \\
-{\mathbf D}_\omega^{-1/2}{\mathbf B}_\omega  & {\mathbf D}_\omega^{1/2}
\end{array}
\right]  \nonumber \\
&&\times
\left[
\begin{array}{cc}
\left(m^*{\mathbf D}_\omega\right)^{1/2}{\mathbf N}^T & {\mathbf 0} \\
  {\mathbf 0}    & \left(m^*{\mathbf D}_\omega\right)^{-1/2}{\mathbf N}^T
\end{array}
\right] \;, \nonumber
\end{eqnarray}
\noindent may be verified to be canonical, {\it i.e.} it satisfies: 
\begin{eqnarray}
{\mathbf C}^T {\mathbf \Sigma}{\mathbf C}={\mathbf \Sigma} \;\;,\;\;\;
{\mathbf \Sigma} = 
\left[
\begin{array}{cc} 
{\mathbf 0} & {\mathbf I}    \\ 
-{\mathbf I}& {\mathbf 0}
\end{array} \right] \;. \nonumber
\end{eqnarray}
\noindent ${\mathbf C}$ therefore transforms the Hamiltonian into 
uncoupled normal modes as required (Eqs. 4 and 5).  

\section{Photoluminescence formula from time dependent 
perturbation theory} 
In this section we {postulate } a time independent Hamiltonian 
${\cal H}^a$ which, on the 
application of first-order time dependent perturbation theory, yields the 
PL formula used (Eqs. $16$ and $11$). (In this section 
all bold faced vectors have three components ($x,y$ and $z$) whereas 
others have only two ($x$ and $y$): ${\mathbf \vec v}=(\vec v,v_z)$.)

The Hamiltonian ${\cal H}^a$ may be assumed to be an approximation to another
Hamiltonian ${\cal H}$ that has been postulated previously \cite{Ct} to 
include the 
(first quantized) degrees of freedom for (spinless) electrons 
(corresponding here to 
unpinned electrons, $N_f$ in number) 
with the electromagnetic fields considered to be the sum of two 
fields - internal fields that are due to the electronic degrees of freedom 
and external fields (here the uniform magnetic field in the $z$ direction and 
the electric fields due to pinned electrons and any required 
confinement potentials) that are independently 
specified. The internal fields are considered dynamical -  their energy 
therefore 
included in ${\cal H}$. These fields are represented through a four vector 
potential $({\mathbf \vec A}^p,\phi^p)$ in the 
Coulomb gauge: ${\mathbf {\vec \nabla} \cdot {\vec A} }^p=0$. It may be seen 
that in this gauge ${\mathbf \vec A}^p$ is dynamical 
(and therefore quantized)
whereas $\phi^p$ is not - it is completely determined  by the 
electronic degrees of freedom as the total Coulomb potential.  Its contribution 
to the field energy results is the Coulomb interaction between the electrons.
The external fields are considered non-dynamical and therefore do not 
contribute to the field energy. They are specified through another four 
vector potential $({\mathbf \vec A},{\mathbf \phi})$ that is a function 
of the electronic degrees of freedom and whose gauge is 
chosen independently. Both the types of 
fields are minimally coupled to all the electronic degrees of freedom. This 
gives: 
\begin{eqnarray}
{\cal H}={\cal H}_0 + {\cal H}_I \;\;\;,\;\;\;{\cal H}_0={\cal H}_0^{e}
+{\cal H}_0^{p} \;,  
\end{eqnarray}
\noindent where
\begin{eqnarray}
{\cal H}_0^e&=&\sum_{k}\left[ \frac{\displaystyle 1}{\displaystyle 2m}
\left[ {\mathbf \vec p}_k - \frac{\displaystyle e}{\displaystyle c}
 {\mathbf \vec A}({\mathbf \vec r}_k) \right]^2 + e\phi({\mathbf \vec r}_k)
+\sum_{l>k}\frac{\displaystyle e^2}{\displaystyle |{\mathbf \vec r}_k - 
{\mathbf \vec r}_l|} \right] \;, \nonumber \\
{\cal H}_0^p&=&\sum_{\displaystyle {\mathbf \vec k},r} 
\hbar\omega_{\mathbf \vec k}
a^\dagger_{{\mathbf \vec k},r}
a_{{\mathbf \vec k},r} \;, \nonumber \\ 
{\cal H}_I&=&\sum_{k} \left[
  \frac{\displaystyle -e}{\displaystyle mc}{\mathbf \vec A}^p
({\mathbf \vec r}_k)\cdot 
\left[{\mathbf \vec p}_k - \frac{\displaystyle e}{\displaystyle c}
 {\mathbf \vec A}({\mathbf \vec r}_k) \right]
+ \frac{\displaystyle e^2}{\displaystyle 2mc^2}
\left[ {\mathbf \vec A}^p({\mathbf \vec r}_k) \right]^2 \right] \;, \nonumber  
\end{eqnarray}
\noindent with 
\begin{eqnarray}
{\mathbf \vec A}^p({\mathbf \vec r})&=&\sqrt{\frac{\displaystyle \hbar c^2}{
\displaystyle 2V\omega_{\vec k}} }
\sum_{{\mathbf \vec k},r} \hat{\epsilon}_{{\mathbf \vec k},r} \left[
a_{{\mathbf\vec k},r}e^{\displaystyle i{\mathbf{\vec k}}\cdot{\mathbf {\vec r}}}
+a^\dagger_{{\mathbf \vec k},r}e^{\displaystyle -i{\mathbf{\vec k}}\cdot
{\mathbf{\vec r}}} \right]  \nonumber \\
\omega_{\mathbf \vec k}&=&c|{\mathbf \vec k}|\; ; \;\hat{\epsilon}_
{{\mathbf \vec k},r}\cdot\hat{\epsilon}_{{\mathbf \vec k},s}=
\delta_{r,s}\;;\;\hat{\epsilon}_{{\mathbf \vec k},r}\cdot{\mathbf \vec k}=0
\;;\;r,s=1,2 \;. \nonumber \\
&&\left\{ a_{{\mathbf\vec k},r},a^\dagger_{{\mathbf \vec k}^\prime,s}
\right\} = \delta_{{\mathbf \vec k},{\mathbf \vec k}^\prime}
                    \delta_{r,s} \;\;\;
\left\{ a_{{\mathbf\vec k},r},a_{{\mathbf \vec k}^\prime,s}
\right\} = 0 \;. \nonumber
\end{eqnarray}
\noindent In the above the index $k$ varies from $1$ to $N_f$, ${\mathbf 
\vec k}$ takes on all values such that ${\mathbf \vec A}^p({\mathbf \vec r})$ 
is periodic over cube-like volumes $V$ and curly brackets imply commutation.

We now go over to our case of a two dimensional electron system by first 
identifying $m$ to be the effective mass $m^*$ and introducing the dielectric 
constant $\epsilon_0$ into the electron-electron interactions. The 
interactions are also modified to take into account the periodic boundary 
conditions used (in section II). 
${\mathbf \vec A}$ is chosen to correspond to the external magnetic field in 
the $z$ direction {\it i.e.} ${\mathbf \vec A}({\mathbf \vec r})=
({\vec A}({\vec r}),0)$. We then 
introduce approximations/assumptions into ${\cal H}_0^e$ that result in 
the loss of the indistinguishability of electrons. The first of these involves 
the external potential $\phi$. It is assumed that $e\phi({\mathbf \vec r})
=V_c(z) + V_p({\mathbf \vec r})$ where $V_p({\mathbf \vec r})$ is due to the 
pinned electrons and $V_c(z)$ is assumed to confine the electrons at particular 
values of $z$. We then assume that $V_c(z)$ takes on two different functional 
forms - the first allowing for only one state in which the electron
is confined to the {\it x-y} plane and the second allowing for an additional 
state in which the electron is confined a distance $z_0$ away from the 
{\it x-y} plane (corresponding to the $z$ position of the holes bound to 
the acceptor atoms) . The first functional form is assumed to apply to all 
the electrons except the recombining electron (index $k=c$) for which 
the second functional form is assumed to apply. This distinguishes the 
the recombining electron from the other electrons. We then make the 
harmonic approximation (as described in section II) which distinguishes 
all the electrons and results in the construction of the two Hamiltonians 
${\cal H}^h_i$ and ${\cal H}^h_f$ corresponding respectively to the states in 
which the $z$ degree of freedom of the recombining electron is confined 
to the {\it x-y} plane or at a distance $z_0$ from this plane. 
As this approximation is made only for the {\it x-y} 
degrees of freedom, the $z$ dependence of $V_p({\mathbf \vec r})$ and 
the electron-electron interactions enters only as parameters of the 
harmonically approximated Hamiltonians ${\cal H}^h_{i,f}$. All the above 
approximations/assumptions are encoded in the following declaration for 
the approximated form (${\cal H}_0^{e,a}$) of the Hamiltonian  
${\cal H}_0^e$ in which the
electronic $z$ degrees of freedom are decoupled from the {\it x-y} degrees
of freedom. With the operators written in the space of 
{\it x-y} electronic degrees of freedom $\otimes$ $z$ electronic degrees 
of freedom, ${\cal H}_0^{e,a}$ may be written as: 
\begin{eqnarray}
{\cal H}_0^{e,a}&=&{\cal H}^h_i\otimes P_0+{\cal H}^h_f\otimes P_{z_0}
\nonumber + I\otimes{\cal H}_z \;, \nonumber \\
{\cal H}_z&=& H^\prime(z_c,p_{z_c}) + \sum_{k\neq c} H(z_k,p_{z_k})\;,\nonumber 
\end{eqnarray}
\noindent where, as mentioned previously, $H(z,p_z)$ admits only 
one state localized around $z=0$ whereas $H^\prime(z,p_z)$ admits an 
additional state localized around $z=z_0$. $P_0$ ($P_1$) is the projection 
operator onto the multiparticle eigenfunction for the $z$ degrees of freedom 
of the electrons $|\psi_0^z\rangle$ ($|\psi_{z_0}^z\rangle$) that has the 
recombining electron localized around $z=0$ ($z=z_0$) - with the corresponding 
eigenvalue with respect to  ${\cal H}_z$ being 0 ($-E^b$). $E^b$ can 
be considered to be the electron-hole binding energy that is an undetermined 
parameter in this theory.

The approximate form of ${\cal H}_0$ is now written as:
\begin{eqnarray}
{\cal H}_0^a = {\cal H}_0^{e,a} + {\cal H}_0^p \;, 
\end{eqnarray}
\noindent which may be seen to have the following eigenfunctions that 
constitute an orthonormal basis:
\begin{eqnarray}
\left\{ |\Psi_I\rangle= |\psi_i\rangle \otimes |\psi_0^z\rangle\otimes 
|\psi^p \rangle \;,\; 
|\Psi_F\rangle= |\psi_f\rangle \otimes |\psi_{z_0}^z\rangle\otimes 
|\psi^p \rangle \right \} \;, \nonumber
\end{eqnarray}
\noindent where (as defined in section II) $|\psi_i\rangle$ 
($|\psi_f\rangle$) is an eigenstate of ${\cal H}^h_i$ (${\cal H}^h_f$) 
with eigenvalue $E_i$ ($E_f$) and $|\psi^p \rangle$ 
is an eigenstate of ${\cal H}_0^p$ that has a definite number of photons 
$n_{{\mathbf \vec k},r}$ in each mode $({\mathbf \vec k},r)$ 
($\Rightarrow$ eigenvalue $E^p=\sum_{{\mathbf k},r}\hbar
\omega_{\mathbf \vec k}n_{{\mathbf \vec k},r}$).
Therefore: 
\begin{eqnarray}
{\cal H}_0^a|\Psi_I\rangle = E_I|\Psi_I\rangle \;\;,\;\;E_I = E_i + E^p  \;,
\nonumber \\
{\cal H}_0^a|\Psi_F\rangle = E_F|\Psi_I\rangle \;\;,\;\;E_F = E_f - E^b + E^p
\;.  \nonumber
\end{eqnarray} 
We now approximate ${\cal H}_I$ (Eq. C1) using: 
${\cal A}^p({\mathbf \vec r}_k)\approx{\cal A}^p({\mathbf \vec r}_k^0)$
 where ${\mathbf \vec r}_k^0$ is the average of the 
classical equilibrium positions of electron $k$ before and after the 
recombination process. This is the dipole approximation that may be 
justified  
using input from the experiment \cite{IVK1} that we are attempting to model 
here - the PL peak is 
at a wavelength $\approx 8230\AA$ (with the width being very 
small:$\approx 20 \AA$) which would be much larger than the spread in 
the ${\mathbf r}_k$ as computed using $|\Psi_{I,F}\rangle$ or the change in the
classical equilibrium value of ${\mathbf r}_k$ due to the recombination process.
Therefore:
\begin{eqnarray}
{\cal H}_I^a&=&\frac{\displaystyle -e}{\displaystyle m^*c}\sum_{k} \left[
  {\vec A}^p ({\mathbf \vec r}_k^0)\cdot
\left[{\vec p}_k - \frac{\displaystyle e}{\displaystyle c}
 {\vec A}({\vec r}_k) \right] 
+ {A}^p_z({\mathbf \vec r}_k^0)p_{z_{k}} \right] \nonumber \\
&&+ \frac{\displaystyle e^2}{\displaystyle 2m^*c^2}\sum_{k}
\left[ {\mathbf \vec A}^p({\mathbf \vec r}_k^0) \right]^2
\end{eqnarray}
\noindent where, in the first summation, the {\it x-y} 
components has been separated from the $z$ component of the vectors. 
 
We now construct the approximate form of ${\cal H}$ (Eq. C1)
using Eqs. C2 and C3: 
\begin{eqnarray}
{\cal H}^a = {\cal H}^a_0 + {\cal H}^a_I
\end{eqnarray}
\noindent  ${\cal H}^a$ is the Hamiltonian in this study. 
First order time dependent   
perturbation theory (Fermi's golden rule) may now be applied to compute 
the intensity of photons emitted $I_+(\omega)$ (or absorbed $I_-(\omega)$). 
The computation is now
analogous to calculations of radiative transitions in atoms - here
all the $N_f$ electrons constitute the ``atom''.  The initial states 
are declared to be of the form $|\Psi_I\rangle$ (recombining electron 
localized around $z=0$) and the final states of 
the form $|\Psi_F\rangle$ (recombining electron localized around $z=z_0$.) 
For emission calculations it is assumed that 
that the initial state $|\Psi_I\rangle_+$ has no photons and the corresponding 
final state $|\Psi_F\rangle_+$ has a single photon with a frequency between 
$\omega$ and $\omega+d\omega$ 
whereas for absorption calculations the opposite is assumed. Initial 
states  $|\Psi_I\rangle_\pm$ that satisfy the requisite constraints on the 
photon numbers 
are further assumed to be thermally distributed. The application of 
the Fermi's golden rule now gives: 
\begin{eqnarray}
I_{\pm}(\omega)d\omega =\frac{ \displaystyle 2\pi}{\displaystyle \hbar} 
\frac{\displaystyle \sum_{I}  e^{\displaystyle -\beta E_I} \sum_{F} 
\left| _\pm\langle\Psi_F|{\cal H}^a_I|\Psi_I\rangle _\pm \right|^2 
\delta(E_I - E_F ) } 
{\displaystyle \sum_{I} e^{\displaystyle -\beta E_I} } \;.  \nonumber 
\end{eqnarray}
\noindent It may be seen that only one term ($-eA^p_z({\mathbf \vec r}_c
^0)p_{z_c}/m^*c$) in Eq. C3, for ${\cal H}^a_I$,  
can have a non-zero matrix element between the states $|\Psi_I\rangle_\pm$ 
and $|\Psi_F\rangle_\pm$ since these differ in the $z$ state of the 
recombining electron. No other term in Eq. C3 has any dependence on the 
operators corresponding to the $z$ degree of freedom of the recombining 
electron ($z_c$,$p_{z_c}$). $I_\pm(\omega)$ is therefore given by the 
following expression in which the photon states $|\psi^p\rangle$ do not appear: 
\begin{eqnarray}
I_\pm(\omega)&=&C(\omega) \left\langle \left[ \hat{\epsilon}_{\hat{r},1}^z 
\right]^2+
\left[ \hat{\epsilon}_{\hat{r},2}^z \right]^2 \right\rangle_{\Delta\Omega}
\Delta\Omega {\cal P}(\pm\omega-\hbar^{-1}E^b) \;, \nonumber \\
&&C(\omega) = \frac{\displaystyle |\omega|}{\displaystyle 8\hbar c}
\left[ \frac{\displaystyle e|\langle\psi^z_{z_0}|p_{z_c}|\psi^z_0\rangle|}
{\displaystyle \pi m^*c} \right]^2 \;, 
\end{eqnarray}
\noindent where $\Delta\Omega$ is the solid angle over which photons of 
momentum ${\mathbf \vec k}= (\omega/c){\hat r}$
are detected (${\hat r}$ being  a unit vector), $\langle\;\;\rangle_
{\Delta\Omega}$ is the average over the solid angle, $\hat{\epsilon}^z$ 
is the $z$ component of $\hat{\epsilon}$ and ${\cal P}$ is given 
by Eq. 11 (section II). From the above it can be seen that 
$I_-(\omega)=I_+(-\omega)$. Hence, both emission and absorption intensities 
can be given by $I_+$ with the convention that negative 
$\omega$ corresponds to absorption of photons. An important prediction 
from the above formula is that photons are fully {\it polarized}. 
This is most easily seen by noting that the transition involves motion of 
an electron in the $z$ direction, which can couple only to the electric fields 
with a $\hat{z}$ component. Since the polarizations of the photons in the 
equations 
below C1 may be specified with one polarization in the {\it x-y} plane, this 
photon does not couple to the transition. Thus, the polarization of an emitted 
photon will be in the $\hat{\mathbf k}\times[\hat{z}\times\hat{\mathbf k}]$
direction.  Lastly, using again 
experimental input that the PL peak width is 
much smaller than the average photon frequency $\omega_{av}$, $C(\omega)$ 
may be replaced by $C(\omega_{av})$. This gives:
\begin{eqnarray}
I(\omega) = C(\omega_{av})\langle \left( \hat{\epsilon}_{\hat{r},1}^z \right)^2
\rangle_{\Delta\Omega}\Delta\Omega {\cal P}(\omega - \hbar^{-1}E^b) \;. 
\end{eqnarray}   
\noindent This justifies Eqs. 16 and 11 which have $E^b=0$. It can now 
be seen that the calculated peak in $I(\omega)$ can be shifted along the 
$\omega$ axis to agree 
with the experimentally observed peak position through a suitable choice of 
$E^b$.

\section{Approximations and convergence criteria}

Here we describe  approximations to
${\cal P}(\omega)$ that provides a cutoff to the integral in
Eq. 12 and also discretizes it to a summation.

It must be noted that ${\cal P}(\omega)$ is 
calculated for a finite system and therefore it consists of a series of
delta functions as can be seen in Eq. 11. The summations in this
equation {\it do not} go over to integrals as would be the case for
an infinite system. Therefore, for ${\cal P}(\omega)$ to be a continuous
curve even in the case of a finite system, the delta function in Eq. 11
must be replaced by a function with non-zero width. We have chosen
this
function to be a Gaussian with width $\Gamma_0$, {\it i.e.} 
\begin{eqnarray}
\lefteqn{ {\mathrm with}\;\;\; g(\Gamma,\omega)=\frac{\displaystyle 1}
{\displaystyle \sqrt{2\pi}\Gamma}e^{-\frac{\displaystyle \omega^2}
{\displaystyle 2\Gamma^2} } \;,   }  \nonumber \\
&{\cal P}_a(\omega) = 
\frac{ \displaystyle
\sum_i e^{-\beta E_i} \sum_f |\langle\psi_f|\psi_i\rangle|^2
g(\Gamma_0,\hbar^{-1}(E_i-E_f)-\omega) }
{\displaystyle \sum_i e^{-\beta E_i}} \;. \nonumber
\end{eqnarray}
\noindent The choice of $\Gamma_0$ 
varies with the cases studied - the corresponding values are specified in 
section III. It can be shown that the approximated ${\cal P}(\omega)$ 
(${\cal P}_a(\omega)$) can be obtained from $\tilde {\cal P}(t)$ as 
given by Eq. 13 using: 
\begin{eqnarray}
{\cal P}_a(\omega) = \frac{\displaystyle 1}{\displaystyle 2\pi}
{\displaystyle \int}_{-\infty}^{+\infty} dt e^{\displaystyle -i\omega t}
e^{\frac{\displaystyle -\Gamma_0^2t^2}{\displaystyle 2}}\tilde{\cal P}(t) \;.
\nonumber
\end{eqnarray}
\noindent The integral can therefore be cutoff between the limits $[-T,T]$ 
such that the error in ${\cal P}_a(\omega)$ due to this 
($\delta{\cal P}_a(\omega)$) is a 
small fraction ($f$) of the expected maximum in ${\cal P}_a(\omega)$ 
(peak height).  Assuming that ${\cal P}_a(\omega)$ consists of a {\it single} 
peak that is Gaussian in shape (this is roughly the experimental line shape 
\cite{IVK1}), the expected peak height in terms 
of the estimated (upper bound) peak width $\sigma_{est}^u$ is equal to 
$\tilde{\cal P}(t=0)/\sqrt{2\pi}\sigma_{est} ^u$.
Further it can be shown that $|\tilde {\cal P}(t)|$ is maximum at $t=0$ (=1 
from Eq. 13). Therefore $T$ is chosen such that:
\begin{eqnarray} 
|\delta{\cal P}_a(\omega)|\leq\tilde{\cal P}(0)\times 2{\displaystyle \int}
_{\displaystyle T}^{\displaystyle \infty}e^{\frac{\displaystyle 
-\Gamma_0^2t^2}{\displaystyle 2}}&dt 
\leq\frac{\displaystyle\tilde{\cal P}(0) f}
{\displaystyle \sqrt{2\pi}\sigma_{est}^u} \nonumber \\
\Rightarrow T \;\;| \;\; 
\Gamma(\frac{\displaystyle 1}
{\displaystyle 2},\frac{\displaystyle \Gamma_0^2T^2}{\displaystyle 2})  
\leq \frac{\displaystyle f\sqrt{\pi}\Gamma_0}{\sigma_{est}^u} &
\end{eqnarray}
\noindent (The gamma function $\Gamma(a,b)$ has been defined in Appendix A.) 
$f$ is chosen to be $10^{-4}$ throughout this study. The choice of  
$\sigma_{est}^u$ varies with the cases studied - its values chosen 
conservatively to be larger than the expected variance of $\omega$ in the 
spectrum being computed.
${\cal P}_a(\omega)$ is therefore given by:
\begin{eqnarray}
{\cal P}_a(\omega) = \frac{\displaystyle 1}{\displaystyle 2\pi}
{\displaystyle \int}_{-T}^{+T} dt e^{\displaystyle -i\omega t}
e^{\frac{\displaystyle -\Gamma_0^2t^2}{\displaystyle 2}}\tilde{\cal P}(t) \;,
\nonumber \\
{\mathrm with} \;\;\;\;\; \omega = n\frac{\displaystyle 2\pi}{\displaystyle 2T}
\;,\;\;\;\;\;n\in {\mathrm integers,} \nonumber
\end{eqnarray}
\noindent where the discretization of $\omega$ follows from the finiteness of 
the time domain integral. 

We now present a modification to $\tilde{\cal P}(t)$ that aids in 
approximating  the 
above integral to a summation and also justifies the assumption that (
the modified) ${\cal P}_a(\omega)$ consists of only one peak. 
As mentioned in the previous section, the 
parameter $\lambda_f$ in ${\cal H}_f$ is chosen to be large so as to 
``suppress'' the {\it x-y} degrees of freedom of the recombining electron 
into becoming ``irrelevant'' in ${\cal H}_f$. We have chosen it to be 
$10^3\;e^2n^{1.5}/\epsilon_0$ throughout this study. This makes 
{\it two} normal modes of ${\cal H}_f^h$ have very large frequencies.
These ``$\lambda$ modes'' may be identified as those whose frequency 
has the strongest dependence on the value of $\lambda_f$. It is found 
(in systems with $N_f=9$) that the initial states $|\psi_i\rangle$ {\it do} 
have a significant overlap with final states $|\psi_f\rangle$ whose 
$\lambda$ modes are excited {\it i.e.} then final state $|\psi_f\rangle$ has 
one or more of the $\lambda$ mode phonons. As a result the 
${\cal P}_a(\omega)$ consists of peaks  
that are separated from each other by 
the $\lambda$ mode frequencies with only a small fraction of the total area 
under the first peak closest to $\omega =0$ (total area$=\tilde {\cal P}
(t=0)=1$). All the peaks are centered around points with 
$\omega <0$ since the ground state energy of ${\cal H}_f^h$ is 
greater than that of ${\cal H}_i^h$ (which is primarily due to the zero point 
energy of the $\lambda$ modes in ${\cal H}_f^h$ as well as because 
$\epsilon^m_f>\epsilon^m_i$). In other words, in the absence of an 
additional term in ${\cal H}_f^h$ corresponding to the (negative) binding 
energy of the electron to the hole all the peaks correspond to absorption 
rather than emission of photons (see Appendix C). We presume that 
{\it only the first peak} is relevant to the experiment {\it i.e.} 
we assume that the electron-hole binding energy (which is not determined within 
our model) is sufficient to shift the peaks such that only the first peak comes 
into the positive $\omega$ domain. This identification is motivated by the 
assumption that the final states $|\psi_f\rangle$ with no $\lambda$ mode 
phonons 
best represent the ``true'' state of the recombined electron (with a more 
realistic Hamiltonian) since, of the different $|\psi_f\rangle$ with 
varying number of $\lambda$ mode phonons, it is these states that maximally 
localize the recombined electron. 
                           
Having argued that only the first peak is of interest, the other peaks may 
by removed from  ${\cal P}_a(\omega)$ and thereby reduce its 
width to that of the first peak. This is achieved by exponentially suppressing 
the irrelevant peaks {i.e.} ${\cal P}_a(\omega)$ is modified to 
${\cal P}_{am}(\omega)$ given by: 
\begin{eqnarray}
\lefteqn{ {\cal P}_{am}(\omega)= \left[ {\displaystyle \sum_i} 
e^{-\beta E_i}\right ] ^{-1} \sum_i e^{-\beta E_i} \times } \nonumber \\
&{\displaystyle \sum}_f |\langle\psi_f|\psi_i\rangle|^2
e^{\displaystyle -\gamma(n^\lambda_1+n^\lambda_2)} 
g(\Gamma_0,\hbar^{-1}(E_i-E_f)-\omega) \; , \nonumber
\end{eqnarray}
\noindent where $n^\lambda_1$ and $n^\lambda_2$ are the numbers of the two 
$\lambda$ mode phonons in $|\psi_f\rangle$. It can be shown (using 
arguments similar to those leading to Eqs. 12-15) that its Fourier 
transform $\tilde {\cal P}_{am}(t)$ is given by: 
\begin{eqnarray}
\tilde{\cal P}_{am}(t)&=&e^{\frac{\displaystyle -\Gamma_0^2t^2}
{\displaystyle 2}}
\left(Tr \left[ e^{\displaystyle -\beta {\cal H}_i^h}   \right] \right)^{-1} 
\times   \\ 
&&e^{ \gamma} Tr \left[
e^{\displaystyle -i\hbar^{-1} \acute{\cal H}_f^ht     }
e^{\displaystyle i\hbar^{-1} {\cal H}_i^h(t+i\hbar\beta)}
   \right]  \;, \nonumber 
\end{eqnarray}
\noindent with 
\begin{eqnarray}
\acute {\cal H}_f^h&=&{\cal H}_f^h({\mathbf \Omega}_f
\longrightarrow {\mathbf \Omega}_f - i\frac{\displaystyle \mathbf{D}_\gamma}
{\displaystyle t}) \;, \nonumber \\ 
&=&\frac{\displaystyle \hbar}{\displaystyle 2}
  \left[\tilde{\mathbf a}_f \tilde{\mathbf a}_f^\dagger \right] 
  \left[ {\mathbf \acute{\Omega}}_f \right]
  \left[ \begin{array}{c}
           {\mathbf a}_f \\
           {\mathbf a}_f^\dagger
         \end{array}           \right] + \epsilon^m_f \;,\nonumber \\ 
{\mathbf \acute{\Omega}}_f&=& 
  \left[ \begin{array}{ll}
          {\mathbf 0}     & {\mathbf \Omega}_f -i\frac{\displaystyle
                                     {\mathbf D}_\gamma}{\displaystyle t}  \\
          {\mathbf \Omega}_f -i\frac{\displaystyle
                       {\mathbf D}_\gamma}{\displaystyle t}& {\mathbf 0}
         \end{array}
  \right]  \;,
\end{eqnarray}
\noindent where ${\mathbf D}_\gamma$ is a diagonal matrix with $\gamma$ 
corresponding to the two $\lambda$ modes and zero otherwise. Calculation of 
the traces appearing in Eq. D2 has been outlined in Appendix E.\cite{DM}
Hence ${\cal P}_{am}(\omega)$ given by:
\begin{eqnarray}
{\cal P}_{am}(\omega) = &\frac{\displaystyle 1}{\displaystyle 2\pi}
{\displaystyle \int}_{-T}^{+T} dt e^{\displaystyle -i\omega t}
\tilde{\cal P}_{am}(t)  \;,\\
&\omega = \frac{\displaystyle n\pi}{\displaystyle T} \;,\nonumber 
\end{eqnarray}
\noindent where the cutoff $T$ continues to be given by Eq. D1. 
The value of $\gamma$ is 
chosen to be $10$ throughout this study. A larger value could not be chosen 
since
numerical accuracy (with double precision numbers) of the matrix operations 
needed to compute $\tilde{\cal P}_{am}(t)$ rapidly decreased with increasing 
$\gamma$. At $\gamma=10$, $\tilde{\cal P}_{am}(t)$ could be calculated 
with a relative accuracy of $\sim 10^{-4}$. In a system with $N_f=9$, it 
has been observed that propagation of this error leads to 
${\cal P}_{am}(\omega)$ having a relative accuracy of $\sim 10^{-4}$ even 
when the limits of integration are $[-10T,10T]$ ($T\approx 61.7 
\left[e^2n^{1.5}/\epsilon_0m^*\right ]^{-1/2}$ for this system). 
(Errors are 
computed relative to a quadruple precision calculation.) As this range of 
integration was the largest in this study, we assume that this numerical 
error is negligible in all cases. 

We can now discretize Eq. D4. To do this correctly with a minimal number of 
$\tilde{\cal P}_{am}(t)$ evaluations {\it i.e.} with the discretization time
being the largest possible, the peak of interest must be at the origin. 
Hence, the 
quantity calculated is ${\cal P}_s(\omega)={\cal P}_{am}(\omega+
\omega_p)$, where $\omega_p$ the is expected peak position 
in ${\cal P}_{am}(\omega)$. In a first approximation this is 
roughly $\omega_0=(E^g_i - E^g_f)/\hbar$ ($<0$) where $E^g_{i,f}$ is the 
ground state energy of ${\cal H}_{i,f}^h$. It is found that $\omega_p$ 
is actually lower than this because the peak is dominated by  
the overlap of the ground state of ${\cal H}_{i}^h$ with {\it excited} 
states of ${\cal H}_{f}^h$. $\omega_p$ 
is therefore calculated assuming that it is equal to the expectation 
value of $\omega$ in the distribution ${\cal P}_{am}(\omega)$:
\begin{eqnarray}
{\mathrm with} \;\;\;\; \omega_p &=& \omega_0 + \bar{\omega} \;,\nonumber \\
{\mathrm where}\;\;\;\;  \omega_0 &=& \frac{\displaystyle 
\epsilon^m_i-\epsilon^m_f}{\displaystyle \hbar} + \frac{\displaystyle 
Tr({\mathbf \Omega}_i - {\mathbf \Omega}_f) }{\displaystyle 2} \;, \nonumber \\ 
{\mathrm and}\;\;\;\;  \bar{\omega}&=&\frac{\displaystyle 1}{\displaystyle i} 
\left. \frac{\displaystyle d}{\displaystyle dt}e^{\displaystyle -i\omega_0t}
\tilde{\cal P}_{am}(t) \right|_{\displaystyle t=0} \;, \nonumber \\ 
{\cal P}_s(\omega)&=&\frac{\displaystyle 1}{\displaystyle 2\pi}
{\displaystyle \int}_{-T}^{+T} dt e^{\displaystyle -i(\omega +\omega_p)t}
\tilde{\cal P}_{am}(t)  \;,\\
&&\omega = \frac{\displaystyle n\pi}{\displaystyle T} \;.\nonumber
\end{eqnarray} 
\noindent The above integration can be transformed into a summation without 
introducing any error if ${\cal P}_s(\omega)$ is 
zero for $|\omega|\geq \omega_M$. Since it is observed that 
${\cal P}_s(\omega)$ 
is roughly Gaussian in shape, we assume that this is indeed the case with 
$\omega_M=5\times(3\sigma_\omega)$ where $\sigma_\omega^2$ is the variance of 
$\omega$ in the distribution ${\cal P}_s(\omega)$ which is given by: 
\begin{eqnarray}
\sigma_\omega^2 = \left. -\frac{\displaystyle d^2}{\displaystyle dt^2}  
e^{\displaystyle -i\omega_0 t} \tilde{\cal P}_{am}(t) \right|_{\displaystyle 
t=0} - \bar{\omega}^2 \; .\nonumber  
\end{eqnarray}
\noindent Eq. D4 is therefore discretized at a rate 
$\geq 2\times\omega_M/2\pi$:\cite{NR}    
\begin{eqnarray}
{\mathrm with}\;\;\;\; \delta t=\frac{\displaystyle L}{\displaystyle T},
\;\;\;\; L = {\mathrm Int.}\left( \frac{\displaystyle T\omega_M}
{\displaystyle \pi} \right) \;, \nonumber \\ 
{\cal P}_s(\omega) = \delta t \sum_{n=-L}^{L-1} e^{\displaystyle -i(\omega
+\omega_p)t} \tilde{\cal P}_{am}(t) \;, \\
\omega = \frac{\displaystyle n\pi}{\displaystyle T}\;,\;\;\;\;-L\leq n \leq L-1 
\;. \nonumber
\end{eqnarray}
\noindent $\omega$ now takes on a finite number of values due to the 
discretization of the time domain. We make use of the symmetry 
$\tilde{\cal P}_{am}(-t) = \left[ \tilde{\cal P}_{am}(t) \right ]^*$ and 
avoid computing $\tilde{\cal P}_{am}(-t)$ for negative values of $t$.  
It must be noted that the large value of 
$\omega_M$ in relation 
to $\sigma_\omega$ is {\it not} forced due to integration errors 
($\sigma_\omega$ is itself an overestimate of the
variance due to the peak of interest since there are contributions
from the other exponentially suppressed peaks) but due 
to the presence of a multivalued function (square root) in 
$\tilde{\cal P}_{am}(t)$ whose correct value has been inferred only by 
assuming that its phase does not vary abruptly between neighboring $t$ 
values.  It must also be noted that the derivatives 
needed to calculate $\bar{\omega}$ and $\sigma_\omega^2$ are computed 
numerically. Finally, ${\cal P}_{am}(\omega)$ is calculated using:
\begin{eqnarray}
{\cal P}_{am}(\omega) = {\cal P}_s(\omega - [\omega_p]) \;,
\end{eqnarray} 
\noindent where $[\omega_p]=\omega_p$ in cases where there is no averaging over 
different ${\cal P}_s(\omega)$. However, in cases that demand such an average, 
$[\omega_p]$ is $\omega_p$ approximated to the nearest integral multiple of 
$\frac{\displaystyle \pi}{ \displaystyle T}$ 
with $\sigma^u_{est}$ being chosen conservatively so that $T$ is the 
same for all the cases being averaged over. This error is expected 
to be negligible since it is verified that 
$\frac{\displaystyle \pi}{\displaystyle T}\ll 3\sigma_\omega$ for each of the 
spectra (${\cal P}_{am}(\omega)$) in the average.

It must be noted that the computational scheme guarentees that 
${\cal P}_{am}(\omega) \in $ real numbers since 
$\tilde{\cal P}_{am}(-t)$ has not been computed independently but has been 
equated to $\left[ \tilde{\cal P}_{am}(t)\right]^*$. However it must be that 
${\cal P}_{am}(\omega)\geq 0$. This condition serves as a good test for the 
validity of the approximations. It has been verified that for all our results 
the largest negative value is always a small fraction 
($\sim 10^{-4}$) of the peak value of ${\cal P}_{am}(\omega)$.

Convergence criteria:
In all the cases studied here the only functions needed to compute the potential
energy terms
(and their derivatives) are the gamma functions $\Gamma(a,b)$ (see
Appendix A). As they have been numerically evaluated with relative error
$\stackrel{\displaystyle <}{\displaystyle \sim}10^{-14}$ 
we extend the summations required to compute the potential ($\tilde 
V$ in Eq. 1 as computed using Eq. A7) to achieve the same accuracy.
Therefore the convergence criterion to determine the (classical) equilibrium
configuration {\it i.e.} the upper bound on the root mean squared value of the 
force per electron
at the end of the propagation with ${\cal H}_i$ is chosen to be $10^{-14}$
natural force units ($=e^2n/\epsilon_0$). Since
the parameter $\lambda_f$ in ${\cal H}_f$ is chosen to be $10^3$ times the
natural
scale of the same dimension, the convergence criterion with ${\cal H}_f$ is
weakened, the corresponding bound being $10^{-11}$ natural force units.

\section{Trace computations with  harmonic Hamiltonians} 

In this section we discuss how the trace operations appearing in Eq. D2 are 
evaluated:
\begin{eqnarray}
Z&=&\left[ e^{\displaystyle - \beta {\cal H}_i^h}   \right] \; ,\nonumber \\
G(t)&=&Tr \left[ e^{\displaystyle -i\hbar^{-1} \acute {\cal H}_f^ht}
e^{\displaystyle i\hbar^{-1} {\cal H}_i^h(t+i\hbar\beta)} \right] \; ,\nonumber 
\end{eqnarray}
\noindent with ${\cal H}_i^h$ given by Eq. 7, $\acute{\cal H}_f^h$ given 
by Eq. D3, and the operators in $\acute{\cal H}_f^h$ being 
related to those in ${\cal H}_i^h$ through Eqs. 8, 9 and 10. We mention that 
the exact evaluation of the traces, which we do by adapting theoretical 
techniques developed in Ref. \onlinecite{DM} in the context of coherent state 
properties, is a central feature of our calculations.

$Z$ is the usual partition function for non-interacting bosons. It may 
be evaluated in the basis of eigenstates of ${\cal H}_i^h$ to be:
\begin{eqnarray}
Z&=&Z_0\prod_{j=1}^{N_f}\left(1 - e^{\displaystyle -\beta\hbar\omega_{i,j}}
\right)
\;\;,\;\;Z_0 = e^{\displaystyle -\beta E^g_i} \;, \nonumber \\
{\mathrm with} &&E^g_i = \hbar 
                       \frac{\displaystyle Tr \left[ {\mathbf \Omega}_i\right]}
{\displaystyle 2} +\epsilon_i^m   \;, \nonumber 
\end{eqnarray}
\noindent where $\omega_{i,j}$ is the diagonal element (phonon frequency) 
of the matrix ${\mathbf \Omega}_i$ appearing ${\cal H}_i^h$. 

Theoretical results of Dodonov and Manko \cite{DM}, that use the coherent 
state basis defined by the operators in ${\cal H}_i^h$, are employed
in evaluating $G$. This basis, $|\mbox{\boldmath $\zeta$}\rangle$, 
is  defined by: 
\begin{eqnarray}
|\mbox{\boldmath $\zeta$}\rangle = e^{\displaystyle 
\tilde{\mathbf a}_i^\dagger\mbox{\boldmath $\zeta$}}|{\mathbf 0}_i \rangle \;,\;
{\mathbf a}_{i,j}|{\mathbf 0}_i \rangle=0 \; ; \;j=1\;{\mathrm to}\;N_f\;,
\; \langle{\mathbf 0}_i|{\mathbf 0}_i\rangle = 1  \; . \nonumber 
\end{eqnarray}
\noindent where $|{\mathbf 0}_i\rangle$ can be identified to be the normalized 
vacuum (ground) state of ${\cal H}_i^h$. $\mbox{\boldmath $\zeta$}$
is a column vector of complex numbers of size $N_f$, and 
$\tilde{\mathbf a}_i^\dagger$ ia a row vector of the 
${\mathbf a}_{i,j}^\dagger$ operators. It can be shown that 
the states $|\mbox{\boldmath $\zeta$}\rangle$ are overcomplete and that the 
trace of an arbitrary operator ${\mathbf \hat{O}}$ can be written as:  
\begin{eqnarray}
Tr \left[ {\mathbf \hat{O}} \right] = \int \frac {\displaystyle d^{ N_f} 
\mbox{\boldmath $\zeta$} } {\displaystyle \pi^{N_f}} 
e^{-\mbox{\boldmath $\zeta$}^\dagger\mbox{\boldmath $\zeta$}}
\langle \mbox{\boldmath $\zeta$}| {\mathbf \hat{O}} | \mbox{\boldmath $\zeta$} 
\rangle \;. \nonumber
\end{eqnarray}
\noindent $G$ may therefore be evaluated, using additionally the cyclic 
permutation invariance of the trace operation and knowing the action of 
the exponential
 of the number operators (${\mathbf {a}}_{i,j}^\dagger{\mathbf a}_{i,j}$) on 
the coherent states, as:
\begin{eqnarray}
G(t)&=&Z_0e^{\displaystyle it\hbar^{-1}E^g_i} 
\int \frac{\displaystyle d^{ N_f} \mbox{\boldmath $\zeta$} } {\pi^{N_f}}
e^{-\mbox{\boldmath $\zeta$}^\dagger\mbox{\boldmath $\zeta$} } 
\langle \mbox{\boldmath $\mu$}|{\cal U} | \mbox{\boldmath $\nu$} \rangle \;,\\ 
{\cal U}&=&e^{\displaystyle -i\hbar^{-1} \acute {\cal H}_f^ht} \;\;,\;\;
\mbox{\boldmath $\mu$}={\mathbf E}^* \mbox{\boldmath $\zeta$} \;\;,\;\; 
\mbox{\boldmath $\nu$}={\mathbf E} \mbox{\boldmath $\zeta$} \;, \nonumber \\
&&{\mathrm with}\;\;\;{\mathbf E}=e^{\displaystyle (-it-\beta\hbar)
\left[{\mathbf \Omega}_i\right]/2} \;. \nonumber
\end{eqnarray}
\noindent Therefore, matrix elements of the ``propagator'' ${\cal U}$ 
corresponding to $\acute {\cal H}_f^h$ 
in the basis $|\mbox{\boldmath $\zeta$}\rangle$ must be evaluated. 
It must be noted that this 
point of view is consistent only if the time variable $t$ that appears 
{\it within} $\acute{\cal H}_f^h$ (Eq. D3) is treated as a 
constant parameter. This parameter is therefore held fixed at $t_0\neq0$ and 
then 
the matrix element is evaluated for all $t$. Of interest would be the value 
when $t=t_0$. (It is assumed that the $t_0=0$ value is equal to the limiting 
value as $t_0\rightarrow 0$.) 
In other words, $\acute {\cal H}_f^h$ is viewed as a time 
independent Hamiltonian. Further, using Eqs. 8, 9 and 10 it can be seen that 
$\acute {\cal H}_f^h$ is a second order polynomial in the operators 
${\mathbf a}_{i,j}$ and ${\mathbf {a}}_{i,j}^\dagger$. The required 
matrix elements for the most generic time dependent Hamiltonian that is of the 
same form has been computed by Dodonov and Manko \cite{DM} in a totally 
different context - we briefly review  
their solution and then specialize to our case.   

The matrix elements of the propagator corresponding to the following 
Hamiltonian are evaluated in the coherent state basis:
\begin{eqnarray}
\frac{\displaystyle {\cal H}}{\displaystyle \hbar} =
\left[ \tilde{\mathbf a}\; \tilde{\mathbf a}^\dagger \right]
\left\{
\frac{\displaystyle {\mathbf K}}{\displaystyle 2}
\left[
\begin{array}{c}
{\mathbf a} \\
{\mathbf a}^\dagger
\end{array}
\right] +
\left[ {\mathbf s} \right]
\right\}   + {\mathbf r} \;\;,\;\;\left( {\mathbf K}^T = {\mathbf K}\right) 
\nonumber  
\end{eqnarray} 
\noindent where the matrix $[{\mathbf K}]$, the vector $[{\mathbf s}]$ and the 
number ${\mathbf r}$ are arbitrary complex functions of time ($t$). 
(The subscript $i$ under 
the operators has been dropped.)  The corresponding propagator ${\cal U}$ (and 
its inverse  ${\cal U}^{-1}$) may be defined through its equation 
of motion together with an initial condition: 
\begin{eqnarray} 
i\hbar\frac{\displaystyle d {\cal U}}{\displaystyle dt}={\cal H}{\cal U} 
\;\;&,&\;\;{\cal U}(t=0) = {\cal I}  \\
i\hbar\frac{\displaystyle d {\cal U}^{-1}}{\displaystyle dt}=-{\cal U}^{-1}
{\cal H} \;\;&,&\;\;{\cal U}^{-1}(t=0) = {\cal I} \nonumber
\end{eqnarray}

It can be shown that an invertable operator can be defined, up to a 
multiplicative constant, by its action on {\it all} the canonically conjugate 
operators, {\it i.e.} if two invertable operators ${\cal U}_x$ and 
${\cal U}_y$ satisfy
\begin{eqnarray}
\left[
\begin{array}{c}
{\mathbf b} \\
{\mathbf \bar{b}}
\end{array}
\right] = 
{\cal U}_{x,(y)}
\left[
\begin{array}{c}
{\mathbf a} \\
{\mathbf \tilde{a}}
\end{array}
\right]  
{\cal U}_{x,(y)}^{-1}
\end{eqnarray} 
\noindent then it must be that ${\cal U}_x=c{\cal U}_y$, where $c$ is a (non 
zero)  c-number.
This is now used to evaluate the propagator 
${\cal U}$ (or equivalently its matrix elements).
The multiplicative freedom in ${\cal U}$ is finally lifted using its
equation of motion together with the initial condition (Eq. E1).  
Note that the elements of
${\mathbf \bar{b}}$ need not be the Hermitian conjugates of those in
${\mathbf b}$ as the operator ${\cal U}$ need not be
unitary (for it could correspond to a non-Hermitian Hamiltonian).

The above procedure to determine ${\cal U}$ can be implemented only if the 
operators ${\mathbf b}$ and ${\mathbf \bar{b}}$ are known. They can be  
determined from their equation of motion together with the initial 
condition which are obtained using Eqs. E2 and E3:
\begin{eqnarray}
i\hbar\frac{\displaystyle d }{\displaystyle dt} 
\left[
\begin{array}{c}
{\mathbf b} \\
{\mathbf \bar{b}}
\end{array}
\right]
=\left\{  {\cal H}\;,\;
\left[
\begin{array}{c}
{\mathbf b} \\
{\mathbf \bar{b}}
\end{array}
\right]
\right \}
\;\;,\;\;
\left[
\begin{array}{c}
{\mathbf b} \\
{\mathbf \bar{b}}
\end{array}
\right] (t=0) = 
\left[
\begin{array}{c}
{\mathbf a} \\
{\mathbf a}^\dagger
\end{array}
\right] 
\end{eqnarray}
\noindent where the curly brackets now represent the commutator. The above 
equations are solved by making the following {\it anzatz} which is motivated 
by ${\cal H}$ being a second order polynomial in the operators ${\mathbf a}_j$
and ${\mathbf a}_j^\dagger$:
\begin{eqnarray}
\left[
\begin{array}{c}
{\mathbf b} \\
{\mathbf \bar{b}}
\end{array}
\right] =
\left[ {\mathbf \Lambda} \right]
\left[
\begin{array}{c}
{\mathbf a} \\
{\mathbf a}^\dagger
\end{array}
\right] +
\left[\mbox{\boldmath $\lambda$} \right]  \;,
\end{eqnarray}
\noindent where ${\mathbf \Lambda}$ ($\mbox{\boldmath $\lambda$}$) is a 
c-number matrix (vector) function of $t$. Substituting the above into Eq. E4 
it can be seen that the 
anzatz is successful as it leads to the following differential 
equations with appropriate initial conditions (the linear independence of 
the operators ${\mathbf a}_j$, ${\mathbf a}_j^\dagger$ and ${\cal I}$ has been 
used): 
\begin{eqnarray}
i\frac{\displaystyle d {\mathbf \Lambda}}{\displaystyle dt} = 
-{\mathbf \Lambda}{\mathbf \Sigma}{\mathbf K} \;\;&,&\;\;{\mathbf \Lambda}(t=0)=
{\mathbf I} \;\;,\;\;{\mathbf \Sigma} = 
\left[
\begin{array}{cc}
{\mathbf 0}  & {\mathbf I} \\
-{\mathbf I} & {\mathbf 0}
\end{array}
\right] \;, \nonumber \\
i\frac{\displaystyle d \mbox{\boldmath $\lambda$}}{\displaystyle dt} = 
-{\mathbf \Lambda}{\mathbf \Sigma}{\mathbf s} \;\;&,&\;\;
\mbox{\boldmath $\lambda$} (t=0)= {\mathbf 0} \;.
\end{eqnarray}
\noindent The above may be solved for ${\mathbf \Lambda}$ and 
$\mbox{\boldmath $\lambda$}$. An important property of the solution is that
the matrix ${\mathbf \Lambda}$ is canonical for all $t$:
\begin{eqnarray}
{\mathbf \Lambda}^T {\mathbf \Sigma}{\mathbf \Lambda}={\mathbf \Sigma} \;.
\end{eqnarray}

Eq. E3 may now be used to 
evaluate the matrix elements of ${\cal U}$ in 
the basis $|\mbox{\boldmath $\zeta$}\rangle$ after substituting for the 
operators 
$\left[{\mathbf b}\right]$ and $\left[{\mathbf \bar{b}}\right]$ using Eq. 
E5 with ${\mathbf \Lambda}$ and $\mbox{\boldmath $\lambda$}$ being the 
solutions of Eq. E6. The matrix elements of Eq. E3 (after multiplying 
from the right by ${\cal U}$) in this basis reads, with
\begin{eqnarray}
{\mathbf \Lambda} = 
\left[
\begin{array}{cc}
{\mathbf \Lambda}_1 & {\mathbf \Lambda}_2 \\
{\mathbf \Lambda}_3 & {\mathbf \Lambda}_4
\end{array}
\right]\;,\;
\mbox{\boldmath $\lambda$}=
\left[
\begin{array}{c}
\mbox{\boldmath $\lambda$}_1 \\
\mbox{\boldmath $\lambda$}_2
\end{array}
\right]\;,\;
\langle\mbox{\boldmath $\mu$}|{\cal U}(t)|\mbox{\boldmath $\nu$}\rangle
=U(\mbox{\boldmath $\mu^*$},\mbox{\boldmath $\nu$},t) \;,\nonumber 
\end{eqnarray}
\begin{eqnarray}
\left\{
{\mathbf \Lambda}_1\frac{\displaystyle \partial\;\;\;}{\displaystyle \partial
\mbox{\boldmath $\mu$}^\dagger }
+{\mathbf \Lambda}_2\mbox{\boldmath $\mu^*$} 
+\mbox{\boldmath $\lambda$}_1
\right\}
U(\mbox{\boldmath $\mu^*$},\mbox{\boldmath $\nu$},t)&=&  
\mbox{\boldmath $\nu$} 
U(\mbox{\boldmath $\mu^*$},\mbox{\boldmath $\nu$},t) \;, \nonumber \\
\left\{
{\mathbf \Lambda}_3\frac{\displaystyle \partial\;\;\;}{\displaystyle \partial
\mbox{\boldmath $\mu$} ^\dagger}
+{\mathbf \Lambda}_4\mbox{\boldmath $\mu^*$} +\mbox{\boldmath $\lambda$}_2
\right\}
U(\mbox{\boldmath $\mu^*$},\mbox{\boldmath $\nu$},t)&=&
\frac{\displaystyle \partial\;\;\;}{\displaystyle \partial 
\mbox{\boldmath $\nu$}^T}
U(\mbox{\boldmath $\mu^*$},\mbox{\boldmath $\nu$},t) \;. \nonumber
\end{eqnarray}
\noindent The above can be solved (assuming the existence of 
${\mathbf \Lambda}_1^{-1}$ and using Eq. E7) for the 
$\mbox{\boldmath $\mu^*$}$ and $\mbox{\boldmath $\nu$}$ dependence of 
$U(\mbox{\boldmath $\mu^*$},\mbox{\boldmath $\nu$},t)$.
The unsolved time dependence appears as an arbitrary multiplicative 
integration constant: 
\begin{eqnarray}
{\mathrm ln}\left[U(\mbox{\boldmath $\mu^*$},\mbox{\boldmath $\nu$},t)\right]=  
\frac{\displaystyle 1}{\displaystyle 2}
\left[ 
-\mbox{\boldmath $\mu$}^\dagger {\mathbf \Lambda}_1^{-1}
{\mathbf \Lambda}_2\mbox{\boldmath $\mu^*$} + 
\mbox{\boldmath $\nu$}^T{\mathbf \Lambda}_3
{\mathbf \Lambda}_1^{-1}\mbox{\boldmath $\nu$}
\right] \nonumber \\
+\mbox{\boldmath $\mu$}^\dagger{\mathbf \Lambda}_1^{-1} 
\left[\mbox{\boldmath $\nu$}-\mbox{\boldmath $\lambda$}_1\right]
+\mbox{\boldmath $\nu$}^T \left[ \mbox{\boldmath $\lambda$}_2 - 
{\mathbf \Lambda}_3{\mathbf \Lambda}_1^{-1}\mbox{\boldmath $\lambda$}_1 \right]
+f(t) \;. 
\end{eqnarray}
\noindent Using the above and evaluating the matrix elements of Eq. E2 
(for ${\cal U}$), it may be seen (using Eqs. E6 and E7) that Eq. E2 
is indeed satisfied in its $\mbox{\boldmath $\mu^*$}$ and 
$\mbox{\boldmath $\nu$}$ dependences, and it leads to the following equations
determining $f$: 
\begin{eqnarray}
\frac{\displaystyle df}{\displaystyle dt} = 
\frac{\displaystyle 1}{\displaystyle 2}\frac{\displaystyle d\;}
{\displaystyle dt}
\left[ 
\mbox{\boldmath $\lambda$}_1^T{\mathbf \Lambda}_3
{\mathbf  \Lambda}_1^{-1}\mbox{\boldmath $\lambda$}_1
\right]  \;, &\\
-\left[ 
i{\mathbf r} +\frac{\displaystyle 1}{\displaystyle 2} Tr 
\right[ {\mathbf \Lambda}_1^{-1} \frac{\displaystyle d {\mathbf \Lambda}_1}
{dt\;\;} \left] +  \frac{\displaystyle d \mbox{\boldmath $\lambda$}_2^T}
{dt\;\;}\mbox{\boldmath $\lambda$}_1
\right] &\;\;,\;\; f(t=0) = 0 \;. \nonumber
\end{eqnarray}
\noindent Hence, as argued previously, Eq. E3 leaves only a c-number 
multiplicative 
freedom in ${\cal U}$ which is lifted by using Eq. E2. This completes 
the generic solution of the matrix elements of interest. 

We now specialize to our case to evaluate $G$ (Eq. E4) by identifying 
${\cal H}$ to be $\acute{\cal H}^h_f$. This implies (using Eqs. 8, 9, 10 and 
D3): 
\begin{eqnarray}
{\mathbf K} = {\mathbf T}^T\acute{\mathbf {\Omega}}_f{\mathbf T}\;,\;\;
{\mathbf s} = {\mathbf T}^T \acute{\mathbf {\Omega}}_f {\mathbf w} \;, \;\;
{\mathbf r} = \frac{\displaystyle {\mathbf w}^T \acute{\mathbf {\Omega}}_f
{\mathbf w}}{\displaystyle 2} + \hbar^{-1}\epsilon_f^m \;. \nonumber
\end{eqnarray}
\noindent The above may now be used to obtain explicit solutions to Eqs. E6
and E9 - as mentioned previously the time $t$ that appears within 
$\acute{\mathbf {\Omega}}_f$ is treated as a constant and then the 
solution is evaluated for $t$ being equal to this constant. This gives, with
\begin{eqnarray}
{\mathbf P}=e^{\displaystyle \left[i{\mathbf \Omega}_ft+{\mathbf D}_\gamma
\right]} \;\;,\;\; 
{\mathbf \acute{P}} = e^{\displaystyle \left[-i{\mathbf \Omega}_ft-{\mathbf D}
_\gamma \right]} \;\;, \nonumber
\end{eqnarray}
\noindent and knowing (using Eqs. 9 and $B4$) that ${\mathbf T}$ is 
canonical, {\it i.e.} ${\mathbf T}^T{\mathbf \Sigma}{\mathbf T}
={\mathbf \Sigma}$: 
\begin{eqnarray}
{\mathbf \Lambda}_1(t)&=&{\mathbf U}^\dagger {\mathbf P}{\mathbf U}
-{\mathbf V}^T {\mathbf \acute{P}}{\mathbf V}^*  \;,  \nonumber \\ 
{\mathbf \Lambda}_2(t)&=&{\mathbf U}^\dagger{\mathbf P}{\mathbf V}  
-{\mathbf V}^T {\mathbf \acute{P}} {\mathbf U}^*  \;,  \nonumber \\
{\mathbf \Lambda}_3(t)&=& {\mathbf U}^T{\mathbf \acute{P}}{\mathbf V}^*  
-{\mathbf V}^\dagger{\mathbf P}{\mathbf U} \;,     \nonumber \\ 
{\mathbf \Lambda}_4(t)&=& {\mathbf U}^T {\mathbf \acute{P}}{\mathbf U}^*  
-{\mathbf V}^\dagger {\mathbf P}{\mathbf V} \;,   \nonumber \\ 
\mbox{\boldmath $\lambda$}_1(t)&=&{\mathbf U}^\dagger 
\left[{\mathbf P}-{\mathbf I} \right]{\mathbf \Delta}
-{\mathbf V}^T \left[{\mathbf \acute{P}}-{\mathbf I}\right]{\mathbf \Delta}^*
\;, \\
\mbox{\boldmath $\lambda$}_2(t)&=&{\mathbf U}^T \left[ {\mathbf \acute{P}}-
{\mathbf I} \right]{\mathbf \Delta}^* -{\mathbf V}^\dagger 
\left[{\mathbf P}-{\mathbf I}\right] {\mathbf \Delta} \;,  \nonumber \\
f(t)&=& \frac{\displaystyle 1}{\displaystyle 2}
\left[
\mbox{\boldmath $\lambda$}_1^T{\mathbf \Lambda}_3
{\mathbf \Lambda}_1^{-1}\mbox{\boldmath $\lambda$}_1
- {\mathrm ln}\left({\mathrm det}{\mathbf \Lambda}_1\right)
\right]  \nonumber \\  
&&+\frac{\displaystyle 1}{\displaystyle 4} 
\left[ \mbox{\boldmath $\lambda$}_1^\dagger\mbox{\boldmath $\lambda$}_1 
-\mbox{\boldmath $\lambda$}_2^\dagger\mbox{\boldmath $\lambda$}_2  
-2\mbox{\boldmath $\lambda$}_2^\dagger\mbox{\boldmath $\lambda$}_1^* \right] 
\nonumber \\
&&-{\mathbf \Delta}^\dagger {\mathrm Sinh}\left[{\mathbf D}_\gamma\right]
\left\{ 2{\mathrm Cos}\left[\Omega_ft\right]-{\mathrm Cosh}\left[{\mathbf D}
_\gamma\right] \right\} {\mathbf \Delta} \nonumber \\
&&-i{\mathbf \Delta}^\dagger {\mathrm Cosh}\left[{\mathbf D}_\gamma\right]
{\mathrm Sin}\left[\Omega_ft\right]{\mathbf \Delta} 
-i\hbar^{-1}\epsilon_f^mt \; . \nonumber
\end{eqnarray}

$G(t)$ (Eq. E1) is now determined as the matrix element 
$\langle \mbox{\boldmath $\mu$}|{\cal U}| \mbox{\boldmath $\nu$} \rangle
=U(\mbox{\boldmath $\mu^*$},\mbox{\boldmath $\nu$},t)$ 
(using Eq. E10) is given by Eq. E8. (Dodonov and Manko \cite{DM} also 
obtain this matrix element for the special case of ${\mathbf \Omega}_f\propto
{\mathbf  I}$ and ${\mathbf D}_\gamma={\mathbf 0}$) Further, the integral in 
Eq. E1 may be carried out analytically (assuming the positive definiteness 
of the appropriate matrix) to give: 
\begin{eqnarray}
{\mathrm ln}\left[\frac{\displaystyle G(t)}{\displaystyle Z_0}\right]&=& 
i\hbar^{-1}E^g_i+ f(t)  \\
&&+ \frac{\displaystyle {\mathrm ln}\left\{{\mathrm det}
\left[{\mathbf A \Sigma_+}\right]\right\} }{\displaystyle 2}
+ \frac{\displaystyle 1}{\displaystyle 2}{\mathbf F}^T{\mathbf A}^{-1}
{\mathbf F} \; , \nonumber \\  
{\mathrm where}\;\;
{\mathbf  \Sigma_+}&=& \left[
\begin{array}{cc}
{\mathbf 0} & {\mathbf I} \\
{\mathbf I} & {\mathbf 0}
\end{array}
\right] \;\;,\;\;
{\mathbf A }={\mathbf  \Sigma_+} - \acute{\mathbf {E}}{\mathbf J}
\acute{\mathbf {E}}\;\;,\;\;
{\mathbf F} =\acute{\mathbf E}{\mathbf l} \;, \nonumber \\  
{\mathrm with}\;\; {\mathbf J}&=& 
\left[
\begin{array}{cc}
{\mathbf \Lambda}_1^{-1}{\mathbf \Lambda}_2 & {\mathbf \Lambda}_1^{-1} \\
\left[{\mathbf\Lambda}_1^{-1}\right]^T&{\mathbf\Lambda}_3{\mathbf\Lambda}_1^{-1}
\end{array}
\right] \;\;,\;\;
\acute{\mathbf E}=
\left[
\begin{array}{cc}
{\mathbf E} & {\mathbf 0} \\
{\mathbf 0} & {\mathbf E}
\end{array}
\right] \;, \nonumber \\
{\mathrm and} \;\;\; {\mathbf l}& = & 
\left[
\begin{array}{c}
-\mbox{\boldmath $\lambda$}_1  \\
\mbox{\boldmath $\lambda$}_2 - {\mathbf \Lambda}_3{\mathbf \Lambda}_1^{-1}
\mbox{\boldmath $\lambda$}_1  
\end{array} \right] \; . \nonumber
\end{eqnarray}
\noindent This completes the analytical determination of the required trace 
operations.

\end{document}